\documentclass[traditabstract]{aa}
\usepackage{natbib}
\usepackage{graphicx}
\usepackage{txfonts}
\usepackage{amssymb}
\usepackage[dvipsnames]{color} 
\usepackage{color}

\begin{document}

\newcommand{\3}{\ss}
\newcommand{\n}{\noindent}
\newcommand{\eps}{\varepsilon}
\newcommand{\be}{\begin{equation}}
\newcommand{\ee}{\end{equation}}
\newcommand{\bl}[1]{\mbox{\boldmath$ #1 $}}
\newcommand{\comment}[1]{\textcolor{red}{ #1}\par}

\def\ba{\begin{eqnarray}}
\def\ea{\end{eqnarray}}
\def\de{\partial}
\def\msun{M_\odot}
\def\div{\nabla\cdot}
\def\grad{\nabla}
\def\rot{\nabla\times}
\def\ltsima{$\; \buildrel < \over \sim \;$}
\def\simlt{\lower.5ex\hbox{\ltsima}}
\def\gtsima{$\; \buildrel > \over \sim \;$}
\def\simgt{\lower.5ex\hbox{\gtsima}}

\authorrunning{Vorobyov et al.}
\titlerunning{Self-gravitating equilibrium models}
\title{Self-gravitating equilibrium models of dwarf galaxies and the minimum mass for star
formation}

\author{Eduard I. Vorobyov, \inst{1,2} Simone Recchi, \inst{1} and Gerhard Hensler, \inst{1} } 
   
   \institute{University of Vienna, Institute of Astrophysics, Vienna, 1180, Austria
\and
Research Institute of Physics, Southern Federal University, Rostov-on-Don, 344090 Russia}

   \date{Accepted by A\&A on May 4, 2012}

\abstract{
We construct a series of model galaxies in rotational equilibrium
consisting of gas, stars, and a fixed dark matter (DM) halo and study how these 
equilibrium systems depend on the mass and form of the
DM halo, gas temperature, non-thermal and rotation support against gravity, and also on the redshift of galaxy formation.
For every model galaxy we find the minimum gas mass $M_{\rm g}^{\rm min}$ required to achieve 
a state in which star formation (SF) is allowed according to contemporary SF criteria.
The obtained $M_{\rm g}^{\rm min}$--$M_{\rm DM}$ relations are compared against the 
baryon-to-DM mass relation $M_{\rm b}$--$M_{\rm DM}$ inferred from the $\Lambda$CDM theory and 
WMAP4 data. 
Our aim is to construct realistic initial models of dwarf galaxies (DGs), which take 
into account the gas self-gravity and can be used as a basis to study the dynamical and chemical 
evolution of DGs.
Rotating equilibria are found by solving numerically the steady-state momentum equation for 
the gas component in the combined gravitational potential of gas, stars, and DM halo using 
a forward substitution procedure.
We find that for a given $M_{\rm DM}$ the value of $M_{\rm g}^{\rm min}$ depends 
crucially on the gas temperature $T_{\rm g}$, gas spin parameter $\alpha$, degree of non-thermal
support $\sigma_{\rm eff}$, and somewhat on the redshift for galaxy formation $z_{\rm gf}$. 
Depending on the actual values of $T_{\rm g}$, $\alpha$, $\sigma_{\rm eff}$, and $z_{\rm gf}$, 
model galaxies may have $M_{\rm g}^{\rm min}$ that are either greater or smaller than 
$M_{\rm b}$. Galaxies with $M_{\rm DM}\ga 10^{9}~M_\odot$ are usually characterized 
by $M_{\rm g}^{\rm min}\la M_{\rm b}$, implying that SF in such objects
is a natural outcome as the required gas mass is consistent with what is available 
according to the $\Lambda$CDM theory. On the other hand, models with 
$M_{\rm DM}\la 10^{9}~M_\odot$ are often characterized by $M_{\rm g}^{\rm min}\gg M_{\rm b}$, implying
that they need much more gas than available to achieve a state in which SF is allowed. 
Our modeling suggests that a star-formation-allowed state is more difficult to achieve in 
DM halos with mass $\la 10^9~M_\odot$ than in their upper-mass counterparts, because 
the required gas mass often exceeds both $M_{\rm b}$ and $M_{\rm DM}$. In the framework of
the $\Lambda$CDM theory, this implies the existence of a critical DM halo mass  below which 
the likelihood of star formation and hence the total stellar mass may drop substantially, 
in accordance with the stellar versus DM halo mass relations recently derived from the SDSS 
survey and Millennium Simulations. On the other hand, DGs that do not follow the $\Lambda$CDM
trend are feasible and have recently been identified, which raises questions about 
the universality of the $\Lambda$CDM paradigm. }

\maketitle

\section{Introduction}
The study of equilibrium states of self-gravitating, multi-component
fluids is of considerable interest in astronomy because they serve as
basic models of many astrophysical objects (stars, protoplanetary
disks, galaxies among others).  While it is known (and quite obvious
from simple symmetry considerations) that isolated, non-rotating,
self-gravitating fluids of finite extent must be spherically
symmetric, it has been a formidable endeavour for some of the most
distinguished astronomers and mathematicians of the last three
centuries to discover the figures of equilibrium in the presence of
rotation.   
In particular, the discovery of Jacobi in 1834 that equilibrium
figures of uniformly rotating fluids need not be axisymmetric took the
scientific community by surprise.   
The formidable body of knowledge on (incompressible and uniformly
rotating) equilibrium figures has been filled up, corrected and
consolidated only recently \citep{chan69, tass78}.

Unfortunately, this knowledge has proven to be inadequate for
the study of equilibrium configurations in galaxies mainly for two
reasons: $(i)$ the gas in galaxies is not incompressible, $(ii)$
galaxies are not uniformly rotating.  Fortunately enough,
non-axisymmetric structures of equilibrium are secularly transformed
into axisymmetric figures in realistic (compressible, viscous and
differentially rotating) models of galaxies \citep[e.g.][]{lind92}.  
It is therefore always realistic to assume that the
equilibrium configuration of a galaxy rotating about some axis is
axisymmetric. It is however not always true  
that this figure of equilibrium is an ellipsoid.  Rapidly rotating, compressible gases
characterized by a polytropic equation of state quite naturally
develop a flared structure \citep{bo73,tass78}.  
Flaring gas distributions in some DGs have been inferred 
\citep{obrien10,bane11} but direct observations of flaring gas disks 
are technically very difficult, even if DGs are edge-on \citep{sa79}. 
Nevertheless, moderate flaring have been observed in the Galaxy \citep{kdf91,kal09} 
and in M31 \citep{bb84} and it is thus reasonable to expect 
flaring also in some gas-rich DGs.  Because of the complex geometries 
(and, often, of the rotation curves) of realistic galaxies, it is extremely complex (if
not impossible) to analytically compute figures of equilibrium and one
must resort to numerical methods.

An equilibrium model without gas 
self-gravity suffers from two major drawbacks. First, such models 
cannot in principle be used to infer equilibrium configurations prone to star formation
since the star formation criteria explicitly or implicitly rely on self-gravity as one of the 
key ingredients for star formation. This lack of self-consistency may lead to situations when 
the star formation feedback due to supernova explosions is studied in models that are insusceptible
to star formation in the first place.
Second, neglecting gas self-gravity one runs the risk of building a
gravitationally overstable configuration, which would have never been realized if
self-gravity were taken into account. Such a non-self-gravitating 
configuration would have too much gas compared to the self-gravitating counterpart and 
additional theoretical or empirical criteria are usually invoked to constrain the
total gas mass \citep[see e.g.][]{MF99,VVS08}. Moreover, the
energy release and the corresponding  SF rates  are often set
arbitrarily \citep[e.g.][]{MF99}.

This paper is the first of a series of works dealing with the
dynamical and chemical evolution of gas and stars in DGs
embedded in dark matter (DM) halos.  In the context, 
achieving an initial equilibrium configuration is clearly
necessary in order to study how the onset of an episode of star
formation or of another perturbing phenomenon affects the evolution of
the studied object.  Surprisingly, almost all the papers on this
subject neglect self-gravity of gas and stars
and consider a simplified initial equilibrium configuration,
namely a rotating isothermal gas distribution in hydrostatic
equilibrium with a fixed potential well \citep[a DM halo or a static
distribution of stars; see e.g.][among many
others]{such94,MF99,ss00,rmd01,mbd03,vorob04,sb10}.  

In this work, we solve numerically the steady-state momentum equation
of a multi-component galaxy (made of gas, stars and a DM halo) taking
into account the gravitational acceleration of all these
components.  This task has been attempted only by very few authors
\citep{nj02,harfst,bane11}.  The typical justification for neglecting gas
self-gravity in constructing initial equilibrium models for DGs is 
that ``the gravitational potential of DGs with
$M_g \simlt 10^9$ M$_\odot$ is dominated by the dark matter halo''
\citep{MF99}.  However, it is worth reminding that some authors still
doubt about the presence of massive DM halos around DGs.
For instance, recent observations of the mass-to-light ratios in Virgo Cluster
dwarf ellipticals by \citet{kol11} and in gas-rich DGs by \citet{swat11}
and also studies of structural properties of the Milky Way dwarf spheroidals
\citep[see e.g.][and references therein]{krou10} are substantially
questioning the contribution of DM on small scales.

Moreover (and more
importantly), even if the total mass of a dwarf galaxy is dominated
by a DM halo, within the Holmberg radius most of the galaxy is made of
baryons \citep[see e.g.][]{papa96,swat11}, although some authors
report different claims \citep[e.g.][]{cb89}.  In some numerical works
(which neglect gas self-gravity) it can be clearly noticed that the
assumed DM profile leads to a very low density of the DM component
(much lower than the gas density) in the central region of the
simulated galaxy \citep[for instance in][the central gas density is
$\sim$ 10 times larger than the DM density]{db99}.  Central densities
of the DM halos in DGs can also be inferred from the
observed rotation curves and typical values are quite low;
significantly below 10$^{-24}$ g cm$^{-3}$ \citep{debl08}.  We can
thus conclude that it is very unlikely that gas self-gravity is
negligible in the central parts of gas-rich DGs.  
All the above arguments in favour of gas self-gravity
justify the relevance of the present study.

The plan of the paper is as follows. The basics of the numerical model are
described in Section~\ref{nummodel}. The initial and boundary conditions, as well as the
solution procedure, are summarized in Section~\ref{incond}. The main results are presented
in Sections~\ref{selfg} and \ref{relation}. A comparison of our results with predictions of
the  $\Lambda$CDM theory is given in Section~\ref{LCDM}. The implications for
the evolution of DGs and the model caveats
are discussed in Sections~\ref{discuss} and \ref{caveats}, respectively. 
The main conclusions are summarized in Section~\ref{conclude}.

\section{Numerical model}
\label{nummodel}
An accurate construction of self-gravitating, rotating equilibria
involves solving for the steady-state momentum equations of gas,
stars, and dark matter in their combined gravitational potential.
This is however a difficult and time consuming numerical exercise,
since the density distribution of each component depends on the total
gravitational potential, which in turn depends 
on the spatial distribution of each component. We simplify
our task by making two assumptions.  First, 
we neglect the contribution by the stellar component to the total
gravitational potential throughout most of the paper 
and return to quantify this effect in Section~\ref{stars}.  
Second, we assume that the DM halo has
a fixed form and hence a fixed gravitational potential.
We note that this assumption may break down on timescales much longer than
a galactic orbital period. We plan to investigate the response of the
DM halo in a follow-up study.

The resulting steady-state momentum equation for the gas component in
the total gravitational potential of gas and dark matter takes the
following form.
\begin{equation}
\label{equilib}
{1\over \rho_{\rm g}}{\bl \nabla} P + \left( {\bl v \cdot \nabla}  \right) {\bl v} = {\bl g}_{\rm g} + {\bl g}_{\rm h},
\end{equation} 
where $P=\rho_{\rm g} \sigma_{\rm g}^2$ is the gas pressure, $\sigma_{\rm g}$
is the one-dimensional gas velocity dispersion, $\rho_{\rm g}$ is the gas
volume density, ${\bl v}$ is the gas velocity, and ${\bl g}_{\rm g}$
and ${\bl g}_{\rm h}$ are the gravitational accelerations due to the
gas and DM halo, respectively. The gravitational acceleration ${\bl g}_{\rm g}$
is calculated as ${\bl g}_{\rm g}=-{\bl \nabla}\Phi$, where the gas gravitational 
potential $\Phi$ is obtained via the solution of the Poisson equation
\begin{equation}
\nabla^2\Phi=4 \pi G \rho_{\rm g}.
\label{Possion}
\end{equation}

For rotating equilibria, it is most convenient to expand
equation~(\ref{equilib}) in cylindrical coordinates ($r,z$) with
imposed axial symmetry, i.e., $\partial / \partial \phi=0$. A steady
state solution implies that $v_{\rm z}=0$ and $v_{\rm r}=0$ (but
$v_{\rm \phi}\ne 0$) and the resulting equations are
\begin{eqnarray}
\label{rhoR}
\sigma_{\rm g}^2 {d \ln \rho_{\rm g} \over d r} &=& {v_{\rm \phi}^2 \over r} - 
{d \sigma_{\rm g}^2 \over dr} + g_{\rm g,r} + g_{\rm h,r} \\
\label{rhoZ}
\sigma_{\rm g}^2 {d \ln \rho_{\rm g} \over dz} &=&  - {d \sigma_{\rm g}^2 \over dz} + 
g_{\rm g,z} + g_{\rm h,z}.
\end{eqnarray}
In this study, we assume that the gas temperature is spatially uniform (see Section~\ref{temper}),
which implies that the spatial derivatives of $\sigma_{\rm g}^2$ are zeroed.

Equations~(\ref{rhoR}) and (\ref{rhoZ}) are discretized using a
first-order backward-difference scheme on a cylindrical mesh with $600
\times 600$ grid points assuming the axial and midplane symmetry
around the $z$-axis ($r=0$) and the midplane $z=0$, respectively. The
resulting set of $600^2$ linear equations is solved using a 
forward substitution scheme explained in detail in the Appendix.

\section{Initial conditions}
\label{incond}
\subsection{Dark matter halo setup}
In order to solve equations~(\ref{rhoR}) and (\ref{rhoZ}) for the gas
density $\rho_{\rm g}$, one needs to specify the form of the DM halo.  We take
two distributions that are most often used to fit the rotation curves
of DGs. The first choice is a quasi-isothermal sphere,
which has a flat near-central density distribution and a tail
inversely proportional to the square of the
distance from the galactic center $\varpi=(z^2+r^2)^{1/2}$ and is
described by the following equation
\begin{equation}
\label{qithermal}
\rho_{\rm qis}= {\rho_{\rm 0} \over 1 + (\varpi/r_{\rm 0})^2}.
\end{equation}
The central density $\rho_{\rm 0}$ and the characteristic scale length
of the quasi-isothermal halo can be calculated using the following
relations \citep[e.g.][]{MF99,Silich01}\footnote{We note that \citet{MF99} have
a misprint in their equations which has been corrected in \citet{Silich01}.}
\begin{eqnarray}
\label{scalelength}
r_{\rm 0}&=&0.89\times 10^{-5} \left( {M_{\rm DM} \over M_\odot} \right)^{1/2} H^{1/2} 
\,\, \mathrm{kpc}, \\
\rho_{\rm 0}&=&6.3\times 10^{10} \left(  {M_{\rm DM} \over M_\odot} \right)^{-1/3} H^{-1/3} 
\,\, M_\odot \, \mathrm{kpc}^{-3},
\end{eqnarray}
where $M_{\rm DM}$ is the mass 
of the DM halo contained within the virial radius 
\begin{equation}
\label{virial}
\varpi_{\rm vir}=0.016 \left( {M_{\rm DM} \over M_\odot}  \right)^{1/3}  H^{-2/3} \,\, \mathrm{kpc}.
\end{equation}
We note that for a fixed $H$ (set to 0.65 in the current paper 
for consistency with the work of \citet{MF99}) 
the quasi-isothermal halo is uniquely determined by a
choice of $M_{\rm DM}$.
Finally, the gravitational acceleration of the quasi-isothermal halo
can be written as
\begin{equation}
{\bl g}_{\rm h}= - 4\pi G \rho_0 r_0^3/\varpi^2 \left[ \varpi/r_0 - \arctan\left( \varpi/r_0\right)  
\right] {\bl e}_\ast,
\end{equation}
where ${\bl e}_\ast={\bl \varpi}/\varpi$ is the unit vector.

The second choice for the form of the DM halo is the 
well-known NFW density profile suggested by \citet{NFW}, 
which features a cuspy profile in the inner
regions and a tail inversely proportional to $\varpi^3$
\begin{equation}
\label{cuspyhalo}
\rho_{\rm NFW}= {\rho_{\rm c} \over (\varpi/r_{\rm c}) (1+\varpi/r_{\rm c})^2},
\end{equation}
where $\rho_{\rm c}$ and $r_{\rm c}$ are free parameters. The mass of
the DM halo contained within radius $\varpi$ can be expressed as
\begin{equation}
M_{\rm DM} (\varpi) = {M_{\rm DM} \over f(c)} \left[\ln(1+x) - {x \over 1+x}  \right],
\end{equation}
where $x=\varpi c/\varpi_{\rm vir}$, $f(c)=\ln(1+c) - c/(1+c)$, $c$ is
the concentration parameter, and the virial radius $\varpi_{\rm vir}$
is defined by equation~(\ref{virial}).  The concentration parameter is
determined from the statistics of the $\Lambda$CDM halo concentrations
by \citet{Neto07}
\begin{equation}
\label{cparameter}
c = 4.67 \left(  {M_{\rm DM} \over 10^{14} M_\odot} H^{-1}  \right)^{-0.11}.
\end{equation}
Finally, the gravitational acceleration due to the NFW halo can be
calculated as
\begin{equation}
{\bl g}_{\rm h} = - {G M_{\rm DM}(\varpi) \over \varpi^2} {\bl e}_\ast.
\end{equation}

A DM distribution profile, somewhat intermediate between the
NFW and the quasi-isothermal profiles, has been semi-empirically
introduced by \citet{Burk95}.  It is described by the following
equation:
\begin{equation}
\label{burkert}
\rho_{\rm Burk}= {\rho_{\rm c} \over 
(1+\varpi/r_{\rm c}) [1+(\varpi/r_{\rm c})^2]}.
\end{equation}
\noindent
It is thus a cored profile (as the quasi-isothermal one) which,
in analogy to the NFW profile, declines at large radii as
$\varpi^{-3}$.  Although this profile fits well the rotation curves
of DGs \citep{Burk95, sb00}, we have not taken it into
consideration, because the results adopting this profile are
intermediate between the results with a (cuspy) NFW and a (cored)
quasi-isothermal profile.  As we show in Sect.~\ref{NFWhalo}, 
our results depend very little on the DM profile, hence, for the sake of 
conciseness, we have not considered the Burkert profile.

\subsection{The gas temperature}
\label{temper}
The thermal properties of gas affect the form of the resulting
equilibrium configuration. A fully self-consistent approach requires
solving for the thermal balance equation along with the steady-state
equations~(\ref{rhoR}) and (\ref{rhoZ}). This however entails a
considerable increase in calculation time and, sometimes, results in
poor convergence.

In this study, we take a simpler approach and build equilibrium
configurations for a pre-defined gas temperature.
This approach is justified if the characteristic cooling/heating time
of gas is much shorter than the dynamical time.
The pre-defined gas temperatures are varied in a wide range, starting from 100 K,
typical for the cold atomic clouds, to $\mathrm{a~few} \times 10^4$~K, typical for the warm
diffuse gas.

\subsection{Rotational versus thermal support}
In order to construct rotating equilibria, one needs to specify the
form of the rotation curve.  A usual approach is to set the rotation
velocity of gas $v_{\phi}$ to the circular velocity $v_{\rm circ}$,
thus assuming that the support against gravity comes mainly from
rotation\footnote{Still there will be some support from gas pressure gradients because
circular velocity is not an exact solution of the steady-state equation~(\ref{equilib}) 
with $P\ne0$.} \cite[e.g.][]{MF99}. 
Such an assumption taken blindly may produce flattened gaseous disks with {\it surface} 
densities nearly independent of galactic radius or even increasing outward, which is
unlikely when compared to real systems\footnote{As mentioned in the Introduction, it
is very likely that the gas vertical scale height naturally
increases outward, producing flaring, 
but the vertically integrated gas volume density (i.e., 
surface density) decreases outward.}.

A more general and realistic approach is to assume that part of the
support against gravity comes from pressure gradients and 
to set $v_{\rm \phi}=\alpha v_{\rm circ}$, 
where $\alpha$ is the spin parameter
that determines the relative contribution of rotation to the total
support against gravity.  For $\alpha=1$, the gas disk is almost  totally supported by
rotation, whereas for $\alpha=0$ the disk is thermally supported. The
resulting expression for the rotational velocity of gas is
\begin{equation}
\label{rotation}
v_{\rm \phi} = \alpha \big[r \left( |g_{\rm h,r}(z=0)| + |g_{\rm g,r}(z=0)| \right) \big]^{1/2},
\end{equation}
where the subscript $r$ denotes the radial component of  the gravitational accelerations, 
the latter being calculated in the midplane 
$z=0$\footnote{In practice, we calculate $|g_{\rm h,r}(z=0)|$ and $|g_{\rm g,r}(z=0)|$ 
in the first layer of computational cells lying right above the midplane.}. 
This choice makes the rotation velocity $z$-independent, in concordance 
with the Poincar\'e-Wavre theorem \citep{Lebovitz} for a barotropic gas in rotation 
equilibrium. More realistic rotating equilibria with a negative vertical gradient of $v_{\rm \phi}$
require considering a more general baroclinic gas \citep{Barnabe}, which is out 
of the scope of the present study. Throughout most of the
paper, we use $\alpha=0.9$ \citep{Tomisaka88,ss00} and explore the dependence of our results
on smaller values of $\alpha$ in Section~\ref{spin}.

\subsection{Boundary conditions}
The final step is to specify the values of gas volume density $\rho_{\rm g}$
at the boundaries.  We use a computational box with physical
dimensions of 8.0~kpc along the $r$- and $z$-axes, with the spatial resolution 
of 13.3~pc along each coordinate direction.
Reflecting boundary conditions at the $z$- and $r$-axes are a natural
choice.  In addition, one needs to define the bounding pressure, i.e.,
the values of $\rho_{\rm g}$ and $\sigma_{\rm g}$ at the outer $z$- and
$r$-boundaries, if a galaxy is submerged in a dense and hot
intra-cluster medium. These values however are essentially free
parameters, since they depend on the environment. Hence we decide to
take a different approach and define the value of gas number density
$n_{0,0}$ in the innermost computational cell near the origin
($z=0,r=0)$. This value is kept fixed throughout the iterative
solution procedure (described below) and serves as a ''seed'' density
needed to solve equations~(\ref{rhoR}) and (\ref{rhoZ}).
Increasing/decreasing the value of $n_{0,0}$ would yield more/less
massive gaseous disks of different spatial configuration. 
Thus, at variance with \citet{MF99}, our model galaxies do not have a disk cutoff radius.  
On the other hand, because of that, ours are truly equilibrium
configurations and the disk does not tend to expand into the
intracluster medium as in Mac Low \& Ferrara (see the test problem in the Appendix).

The choice of $\alpha$, $\sigma_{\rm g}$ and $n_{0,0}$, along with
the mean molecular weight $\mu=1.26$ (for a metallicity of 1/100 that of the solar) and 
reflecting boundary conditions at the $z$- and $r$-axes, completes
the initial setup and allow us to calculate equilibrium configurations
of gaseous disks for different shapes and masses of the DM
halo.  The bounding effect of the external 
environment will be addressed in a future study.

\begin{figure}
  \resizebox{\hsize}{!}{\includegraphics{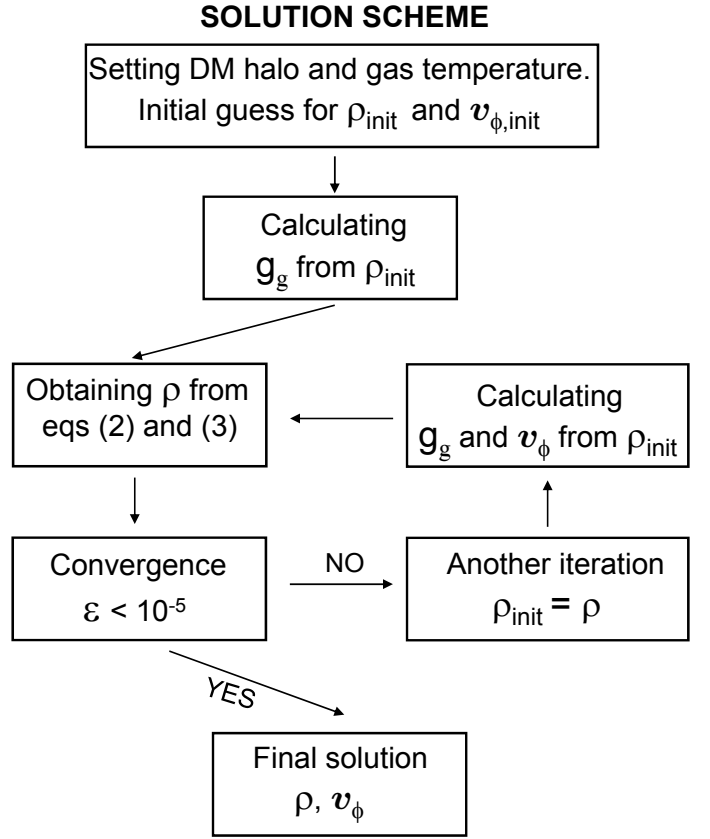}}
      \caption{Schematic representation of the iterative solution procedure to
      calculate rotating, self-gravitating gaseous equilibria in DM halos.}
         \label{fig1}
\end{figure}

\subsection{The solution procedure}
One can notice that, since ${\bl g}_{\rm g}$ is obtained 
through the solution of the Poisson equation~(\ref{Possion}), 
equations~(\ref{rhoR}) and (\ref{rhoZ}) are
transcendental and hence require an
iterative solution procedure described schematically in
Fig.~\ref{fig1}.  The calculation begins with {\it a)} choosing the
DM halo profile (quasi-isothermal sphere or NFW halo), {\it
  b)} calculating the corresponding gravitational acceleration
$\bl{g}_{\rm h}$, {\it c)} fixing the thermal properties of gas, i.e., 
the gas temperature, and {\it d)} making an
initial guess for the gas volume density $\rho_{\rm g,init}$ and
rotational velocity $v_{\rm \phi,init}$.  For the former we usually
choose a spatially uniform distribution with $\rho_{\rm g}=m_{\rm H}\, \mu\, n_{0,0}$ and the
latter is determined from equation~(\ref{rotation}). Finally, the
gravitational acceleration $\bl{g}_{\rm g}$ of the gas
configuration is calculated by solving for the Poisson equation 
using the alternative direction implicit method as described in \cite{Black75} and \cite{SN92}.

The first loop of iterations begins with solving for the steady-state
equations~(\ref{rhoR}) and (\ref{rhoZ}).  The resulting gas density
$\rho_{\rm g}$ is compared against the initial guess $\rho_{\rm g,init}$ for
every computational cell and if the {\it maximum} relative error
$\epsilon$ is larger 
than $10^{-5}$, then the iteration cycle repeats by setting $\rho_{\rm
  g,init}=\rho_{\rm g}$ and calculating new gravitational potential and rotation
curve of the gas disk.  Usually, convergence is achieved after 10--15
iterations, but may sometime diverge signalizing for an inappropriate
initial guess for $n_{0,0}$ or $\alpha$, especially when $n_{0,0}$
is large and the corresponding equilibrium solution near the rotation axis 
is characterized by a narrow, high-density plateau which is difficult
to resolve numerically.

\section{Self-gravitating equilibrium gaseous disks}
\label{selfg}
\subsection{Star formation criteria}
\label{SFcriteria}
We build equilibrium gaseous disks hosted by DM halos of various mass
and shape and determine the minimum gas mass needed to trigger star
formation in these systems.  Three criteria are employed to assess the
feasibility of star formation in our model galaxies.  The first criterion
is based on theoretical considerations of gravitational stability in
self-gravitating systems. We assume that star formation is allowed
if the Toomre $Q$ parameter
\begin{equation}
Q_{\rm T}={\nu \sigma_{\rm g}\over \pi G \Sigma}
\label{Toomre}
\end{equation} 
is smaller than a critical value $Q_{\rm c}$, where $\nu$ is the
epicycle frequency and $\Sigma$ is the gas surface density. The
classical analysis of thin, axisymmetric gaseous disks suggests a
value of $Q_{\rm c}=1.0$ \cite[][]{Toomre64}, but $Q_{\rm c}$ in real systems
 is usually somewhat greater and may depend on many factors
including the galaxy class, the form of the rotation curve, the disk thickness, the
strength of magnetic fields, etc \citep[e.g.][]{Polyachenko97,Ostriker01,Bigiel,Leroy,Dong,Roy09}. 
In this study, we take a conservative value of $Q_{\rm c}=2.0$ and assume 
that our model galaxy is prone to star formation if
$Q_{\rm T}<Q_{\rm c}$ in at least some parts of the gas disk.

For the second star formation criterion, we make use of empirical
studies of star formation in the Local Universe by
\cite{Kennicutt98,Kennicutt08} who compares disk-averaged star
formation (rates per unit area) versus gas surface densities in normal
and starburst galaxies, including DGs.  These studies suggest
the following scaling law (hereafter, the Kennicutt-Schmidt law) 
between the star formation rate
per unit area $\Sigma_{\rm SFR}(M_\odot$~yr$^{-1}$~kpc$^{-2}$) and the gas surface density 
$\Sigma(M_\odot$~pc$^{-2}$)
\begin{equation}
\Sigma_{\rm SFR}= (2.5\pm0.7)\times10^{-4} 
\left( {\Sigma \over \mbox{1~$M_\odot$~pc$^{-2}$} } \right)^{1.4\pm0.15},
\label{KS}
\end{equation}
with a threshold density $\Sigma_{\rm c}$ of the order of 
$5.0~M_\odot$~pc$^{-2}$, below which very rare cases of large-scale 
star formation are detected.
Therefore, we use this value as the second star formation criterion and assume that star
formation can be triggered in our model galaxies if
$\Sigma>\Sigma_{\rm c}$. 

So far, we have used vertically integrated gas densities to assess the model's susceptibility
to star formation. However, star formation recipes may also rely on the critical gas 
{\it volume} density $n_{\rm c}$, as is often done in numerical hydrodynamics simulations. 
Moreover, as discussed in \citet{elm97}, 
a Schmidt law with index 1.5 would be expected for self-gravitating disks, if the SF rate is equal 
to the ratio of the local gas volume density to the free-fall time, all multiplied by some efficiency.
The adopted values of $n_{\rm c}$ vary in wide limits, depending on the numerical resolution but 
most studies use values of the order of 0.1--1.0~cm$^{-3}$ \citep[e.g.][]{Springel03,Schaye08}, though
some authors adopt much higher values \citep[e.g.][]{Tasker11}. 

In this paper, we assume that SF is allowed if there is enough gas mass 
(in the vertical column that fulfils the first two criteria)
with number density $n_{\rm g}$ greater than a fiducial critical value of 
$n_{\rm c}=1.0$~cm$^{-3}$ to allow for a SF event of non-negligible magnitude, i.e., if
\begin{equation}
M_{\rm g}^{\rm SF}\left( n_{\rm g}\ge 1.0~\mathrm{cm}^{-3} \right) > 10^4~M_\odot.
\label{SFmass}
\end{equation}
We choose to set a limit in mass rather than in size because
star formation may be localized to just a few tens of parsec, yet contain enough
gas mass for a star formation event of notable magnitude. 
Finally, a model galaxy is assumed to
be prone to star formation only if all three criteria are satisfied altogether.

\begin{table}
\center
\caption{Quasi-isothermal DM halo parameters}
\label{table1}
\begin{tabular}{|c|c|c|c|}
\hline\hline
$M_{\rm DM}~(M_\odot)$ & $r_0$~(kpc) & $\rho_0$~($M_\odot$~pc$^{-3}$) & $\varpi_{\rm vir}$~(kpc)  \\
\hline
$10^7$ & $2.27\times 10^{-2}$ & 0.337 & 4.6  \\
$10^8$ & $7.2\times 10^{-2}$ & 0.157 & 9.9 \\
$10^9$ & 0.23 & 0.073 & 21.3   \\
$10^{10}$ & 0.72 & 0.034 & 45.9  \\
\hline
\end{tabular}
\end{table}

\subsection{Equilibrium models}
Throughout most of the paper, we use a quasi-isothermal DM
halo with four different masses $M_{\rm
  DM}=10^{7}~M_\odot$, $10^8~M_\odot$, $10^9~M_\odot$, and
$10^{10}~M_\odot$.
The corresponding values for $r_0$, $\rho_0$, and $\varpi$ are listed
in Table~\ref{table1}.  We set the gas spin parameter to $\alpha=0.9$ and
the gas temperature $T_{\rm g}$ to a {\it spatially} uniform  
value that is either independent of the halo
mass ($T_{\rm g}=10^4$~K) or scales with the DM mass as
$T_{\rm g}\propto M_{\rm DM}^{2/3}$, as suggested by the virial
relations.  We consider the effect of varying
rotational support against gravity (i.e., varying $\alpha$) in
Section~\ref{spin}, the effect of a different DM halo
configuration (i.e., the NFW halo) in Section~\ref{NFWhalo}, and the
effect of non-negligible stellar disk in Section~\ref{stars}.

To put things in the physical context, the models
considered here and in Sections~\ref{relation}--\ref{NFWhalo} 
(with a gas distribution in equilibrium with a DM
halo, waiting for the onset of star formation) can be considered as 
progenitors of, e.g., Blue Compact Dwarf galaxies whose stellar populations
are largely dominated by very young stars \citep{Papaderos08}.  In
Section~\ref{stars} we describe objects with a pre-existing disk of
stars which have smoothly accreted gas and have achieved a new
equilibrium configuration (galaxies surrounded by extended gas reservoirs 
are quite common \citep[see e.g.][for the case of I Zw 18]{Zee98}).
Finally, in Section \ref{highzgf} we assume a redshift of galaxy
formation significantly larger than zero.
Therefore, our equilibrium configurations should be treated
as proxies to DGs that have built up their gas mass
reservoir by a quasi-steady accretion or have temporally 
achieved a quasi-steady state after an episode of fast accretion. 

\begin{figure*}
 \centering
  \includegraphics[width=15cm]{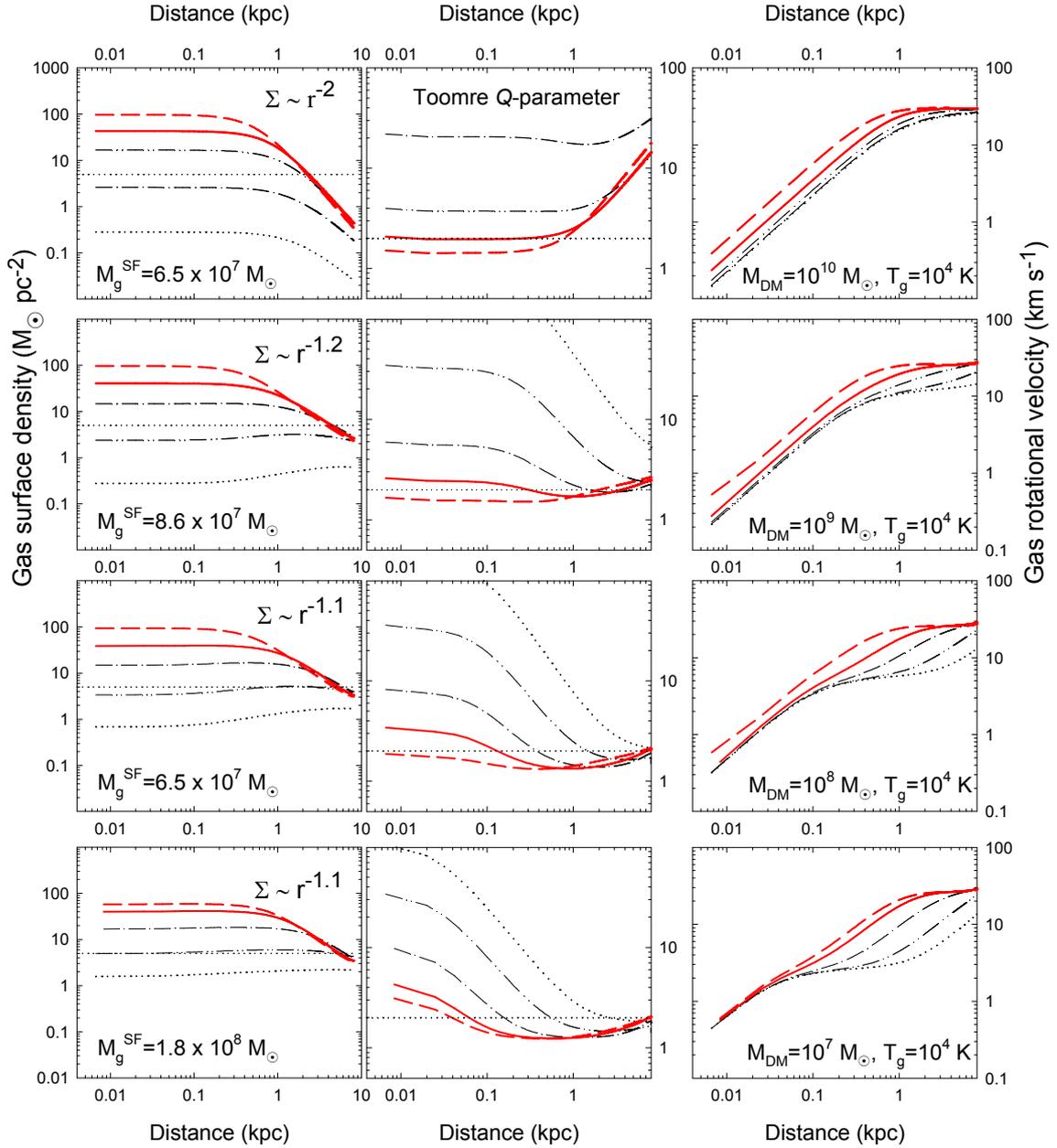}
      \caption{Gas surface densities (left column), Toomre $Q$ parameters (middle column),
      and rotation velocities (right column) of self-gravitating (steady-state) gaseous disk
      hosted by quasi-isothermal DM halos with four masses $M_{\rm DM}$ as indicated in the right column.
      The spatially uniform gas temperature and the spin parameter are set to $T_{\rm g}=10^4$~K 
      and $\alpha=0.9$, respectively. The horizontal dotted lines mark 
the adopted critical surface density 
      for star formation $\Sigma_{\rm c}=5~M_\odot$~pc$^2$ (left column) and the critical Toomre
      parameter $Q_{\rm cr}$ for gravitational stability (middle column).
      For every value of $M_{\rm DM}$, five 
      models with different values of the seed density $n_{0,0}$=25.0, 5.0, 1.0, 0.1, and 0.01 
      cm$^{-3}$ from the uppermost to the lowermost line are considered. 
      Greater $n_{0,0}$ produce models with higher gas surface densities in the inner regions.
      The models that are susceptible to star formation according to the adopted star formation 
      criteria are marked with red thick lines and the gas mass $M_{\rm g}^{\rm SF}$ with number
      density greater than a critical value of 1.0~cm$^{-3}$ is indicated in the left column for models
      plotted with solid red lines.
}
         \label{fig2}
\end{figure*}

Figure~\ref{fig2} presents gas surface densities $\Sigma$ (left
column), Toomre $Q$ parameters (middle column), and gas rotation
velocities $v_\phi$ (right column) for various steady-state gaseous
disks with $M_{\rm DM}$ ranging from $10^{10}~M_\odot$ (top row) to
$10^{7}~M_\odot$ (bottom row).  The spin parameter and the spatially
constant gas temperature are the same for all models and are equal to
0.9 and $10^4$~K, respectively. In the calculation of the
$Q$ parameter and $v_{\rm \phi}$ we use mass-weighting according to
the gas mass contained in every computational cell. The gas surface
density is obtained by integrating $\rho_{\rm g}$ along the $z$-axis.

For each value of $M_{\rm DM}$, we consider five models with different
seed values of the gas number density $n_{0,0}$,
namely, 0.01, 0.1, 1.0, 5.0, and 25~cm$^{-3}$. 
These models are
distinguished in Figure~\ref{fig2} by lines of different style, with
the dotted lines corresponding to $n_{0,0}=0.01$~cm$^{-3}$ 
 and dashed lines to
$n_{0,0}=25$~cm$^{-3}$ (and the other models in between in the order of increasing $n_{0,0}$).
The largest/smallest values of $n_{0,0}$ produce models with highest/lowest 
gas surface densities near the galactic center.

The radial profiles of $\Sigma$ in Figure~\ref{fig2} indicate that models 
with lower values of $M_{\rm DM}$
produce less centrally concentrated gaseous distributions.  
Indeed, models with $M_{\rm DM}=10^{10}~M_\odot$ have a density tail proportional to $r^{-2}$, 
whereas models with $M_{\rm DM}\le10^{8}~M_\odot$ are characterized by $\Sigma\propto r^{-1.1}$.  
This tendency can be explained by the fact that the mass of the gas disk starts 
to systematically exceed that of the DM halo for $M_{\rm DM}\le10^{9}~M_\odot$, 
the effect discussed in more detail in Section~\ref{relation}.
As a result, the shape of the gas disk in models with $M_{\rm DM}\le10^{9}~M_\odot$ 
is mostly determined by self-gravity of the gas, with the resulting distribution approaching
that of a self-gravitating isothermal ellipsoid with the density tail
$\rho_{\rm g} \propto \varpi^{-2}$ or $\Sigma \propto r^{-1}$. We also note
that models with lowest values of $\Sigma$ tend to have surface
density profiles independent of radius.

For a given DM halo mass and equal gas temperature, models with lower values of $\Sigma$ 
have higher values of the $Q$ parameter, as expected. 
It is worth noting that the lowest $Q_{\rm T}$ is often found a few
hundred or even thousand parsecs away from the galactic center. This
behaviour can be understood by analysing the radial dependence of the
epicycle frequency $\nu = (4\Omega^2 + r d\Omega^2/d r)^{1/2}$.  This
quantity is independent of radius $r$ in the inner parts $r\ll r_0$,
where the DM halo and gas densities are nearly constant and
$\Omega\approx \mathrm{const}$.  On the other hand, at $r\gg
r_0$ the epicycle frequency declines with radius because the DM halo and gas densities also
(as a rule) decline with radius.  This implies that $Q_{\rm T}$ is
nearly independent of $r$ in the inner parts but may increase or
decrease in the outer parts depending on the radial profile of the gas
surface density $\Sigma$.  For models with $\Sigma$ nearly independent
of radius, $Q_{\rm T}$ generally decreases at large radii (because $\nu$ also 
decreases but other quantities stay nearly constant), whereas for
models showing a decline in $\Sigma$ at large radii, the corresponding
values of $Q_{\rm T}$ attain a minimum at some several hundred or
thousand parsecs and increases on both sides.  

It is also worth noting that the rotation curves of our model galaxies
either steadily rise or flatten out only at large radii.  This
behaviour is in qualitative agreement with the observed rotation
curves of DGs \citep[see e.g.][]{debl08,Oh08,swat09,swat11}.

The horizontal dotted lines in the left and middle columns of
Fig.~\ref{fig2} mark the critical gas surface density for star
formation $\Sigma_{\rm c}=5~M_\odot$~pc$^2$ and the critical Toomre
parameter $Q_{\rm c}=2.0$. We use these values to determine models
that can allow for star formation, i.e., models for which both criteria 
$\Sigma>\Sigma_{\rm c}$ and $Q_{\rm T}<Q_{\rm c}$ are met 
at least in some parts of the galactic disk, and also the gas mass 
$M_{\rm g}^{\rm SF}$ with number density greater than 1.0~cm$^{-3}$ exceeds
a minimum value of $10^4~M_\odot$ as stipulated by the third SF condition. 
Such ``star-formation-allowed'' models are highlighted with red color in Fig.~\ref{fig2}.

For every model in Figure~\ref{fig2}, we calculate the
total gas mass $M_{\rm g}$ contained within our computational domain,
the latter having a cylindrical shape with radius of 8~kpc and height of 8~kpc on both
sides from the midplane. We use $M_{\rm g}$ as a proxy for the total 
gas mass\footnote{We cannot extend our
computational boundaries to a distance much greater than 8~kpc because of
the need for high spatial resolution to resolve the inner density plateau.},
and estimate the possible missing gas mass using the surface density profiles in Fig.~\ref{fig2}.
For models with $M_{\rm DM}=10^{10}~M_\odot$, $\Sigma(r) \propto r^{-2}$ and 
$M_{\rm g}(r) \propto \ln (r)$ at large radii, 
implying a small correction of order unity for 
a computational box with size three times greater than ours ($24\times24$~kpc).
For models with $M_{\rm DM}\le10^8~M_\odot$, $\Sigma(r) \propto r^{-1.1}$ and
$M_{\rm g}(r) \propto r^{0.9}$, implying a factor of 2.7 increase in
the total gas mass. This means that our estimates are accurate to
within a factor of unity for models with $M_{\rm DM}\ga 10^{9}~M_\odot$, 
while for models with smaller DM halos we may underestimate
the total gas mass by up to a factor of 3. 

\begin{figure}
  \resizebox{\hsize}{!}{\includegraphics{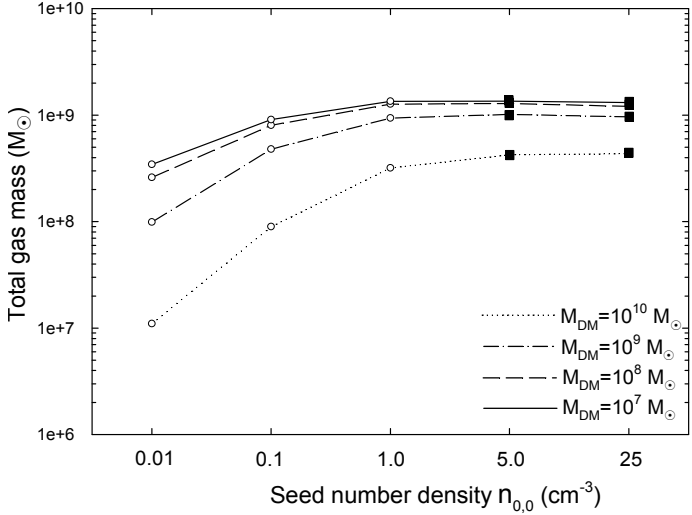}}
      \caption{Total gas masses $M_{\rm g}$ calculated for five models with different seed density values
      $n_{0,0}$. The filled squares mark the models that are susceptible to star formation
      according to the adopted criteria. Lines of different style connect models with equal masses
      of the DM halo.}
         \label{fig3}
\end{figure}

The total gas masses $M_{\rm g}$ for every model are plotted in
Fig.~\ref{fig3}.  
The bottom axis shows the seed number density
$n_{0,0}$ for each model. Lines of different style connect models
with the same mass of the DM halo (e.g., models with $M_{\rm
  DM}=10^9~M_\odot$ are connected with the dash-dotted line).
We distinguish the star-formation-allowed models by the filled squares.  

It is evident that DM halos can host steady-state 
gaseous disks with various masses, but not all gas configurations are prone
to star formation.  There exists a minimum gas mass $M_{\rm g}^{\rm
 min}$ that a DM halo needs to accumulate  
in order to fulfil the star formation criteria.
For instance, for a DM halo with mass $10^{10}~M_\odot$ (dotted line)  
the corresponding minimum gas mass is $4.3\times 10^8~M_\odot$, 
while for a DM halo with mass $10^{8}~M_\odot$ (dashed line) the corresponding 
minimum gas mass is $1.3\times 10^9~M_\odot$. As one can see, $M_{\rm g}^{\rm min}$ depends 
on the mass of the DM halo and may increase as
$M_{\rm DM}$ decreases.  The latter effect is not unexpected -- a DM
halo with smaller mass has  a weaker gravitational potential and, as a consequence, 
a more massive gaseous disk is required  to attain the
critical density for star formation. The gas self-gravity here is a key factor, without 
which such an effect will be absent.

\section{Minimum gas mass for star formation vs. dark matter halo mass}

\label{relation}
In this section, we study in more detail the dependence of the minimum gas mass
for star formation $M_{\rm g}^{\rm min}$ on the mass of the DM
halo, as well as on other properties of galactic systems.
We want to emphasize here that $M_{\rm g}^{\rm min}$ is the total gas mass
of a galaxy in which star formation is allowed according to the adopted SF criteria and
not the gas mass $M_{\rm g}^{\rm SF}$ that fulfils the star formation criterion~(\ref{SFmass}). 
The latter quantity is always smaller than $M_{\rm g}^{\rm min}$ as only part of the gas 
disk is characterized by $n_{\rm g}>n_{\rm c}$.
We choose to concentrate on $M_{\rm g}^{\rm min}$ because we compare this quantity 
to the baryonic mass derived from the $\Lambda$CDM theory.

\begin{figure}
  \resizebox{\hsize}{!}{\includegraphics{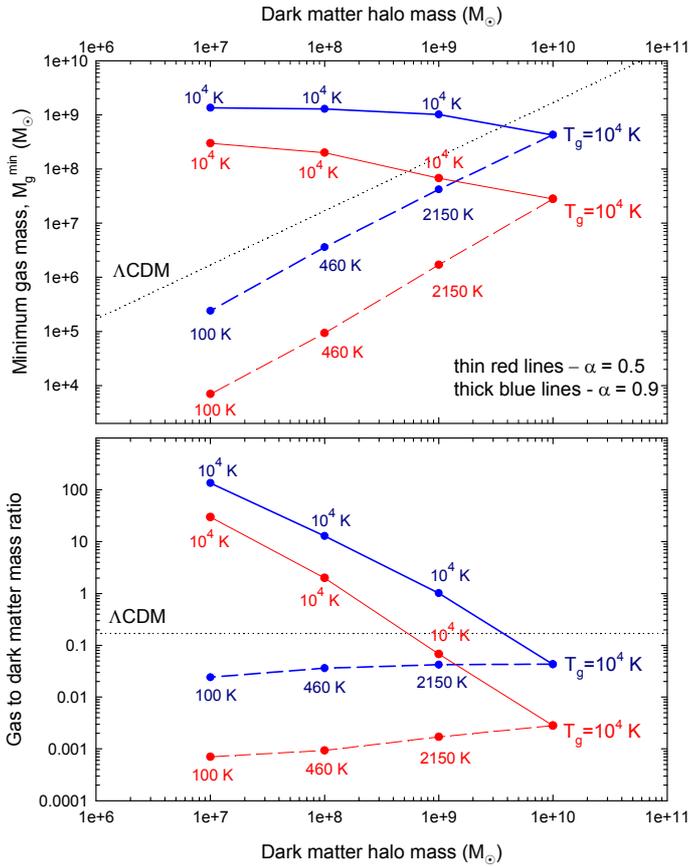}}
      \caption{{\bf Top}. Minimum gas mass for star formation $M_{\rm g}^{\rm min}$ as a function 
      of the DM halo mass $M_{\rm DM}$. The thick blue lines show the data for the reference model (quasi-isothermal
      DM halo, spatially uniform $T_{\rm g}$, and $\alpha=0.9$), while the thin red lines show the effect
      of a smaller spin parameter $\alpha=0.5$. The solid lines (of corresponding color) refer 
      to models with $T_{\rm g}=10^4$~K,
      while the dashed lines show models with gas temperature 
      according to eq.~(\ref{scaling}). Numbers denote the
      corresponding gas temperatures for every model. The dotted line
      gives the available baryonic mass $M_{\rm b}$ as suggested by the $\mathrm{\Lambda CDM}$ 
      theory with 
      $\Omega_{\rm b}/\Omega_{\rm m}=0.17$. {\bf Bottom}. Ratio $M_{\rm g}^{\rm min}/M_{\rm DM}$ versus
      $M_{\rm DM}$ for the same models as in the top panel. }
         \label{fig4}
\end{figure}

Throughout the paper we consider quasi-isothermal DM halos (if not
specified otherwise) described by equations~(\ref{qithermal})-(\ref{virial}).
The top panel in Figure~\ref{fig4} presents the $M_{\rm g}^{\rm
min}$--$M_{\rm DM}$ relation for the four values of the DM halo mass
($10^7~M_\odot$, $10^8~M_\odot$, $10^9~M_\odot$, and $10^{10}~M_\odot$).
In particular, the thick blue solid line shows the data
for the spin parameter $\alpha=0.9$ and gas temperature $T_{\rm g}=10^{4}$~K 
(independent of the DM halo mass), whereas the thick blue dashed line does that for 
$T_{\rm g}\propto M_{\rm DM}^{2/3}$ and the same value of $\alpha$\footnote{   
We remind that in all models the gas temperature is spatially uniform but
its value may or may not depend on $M_{\rm DM}$.}. The latter relation is normalized to
$T_{\rm g}=10^4$~K for $M_{\rm DM}=10^{10}~M_\odot$, 
which yields the following scaling law
\begin{equation}
\label{scaling}
T_{\rm g}=2.15\times10^{-3} M_{\rm DM}^{2/3}.
\end{equation}

The corresponding values of $T_{\rm g}$ are indicated in Fig.~\ref{fig4}
for every pair of data ($M_{\rm g}^{\rm min}, M_{\rm DM}$).
The adopted set of parameters
(quasi-isothermal DM halo, $\alpha=0.9$, and gas temperature either dependent on or
independent of $M_{\rm DM}$) is denoted hereafter as the reference  model. 
The bottom panel in Figure~\ref{fig4} also shows the ratio
$\xi=M_{\rm g}^{\rm min}/M_{\rm DM}$ versus $M_{\rm DM}$.

The dotted line in Figure~\ref{fig4} presents the baryonic mass
$M_{\rm b}$ for a given DM halo mass as expected from the
$\mathrm{\Lambda CDM}$ theory and WMAP4 data \citep{sper03}
with $\Omega_{\rm b}/\Omega_{\rm m}=0.17$. If we treat $M_{\rm b}$ 
as an {\it upper} limit of the available gas mass, it becomes evident 
that galaxies with $M_{\rm DM}\le 10^{9}$ and $T_{\rm g}=10^{4}$~K require
much more gas than available to achieve a state in which star formation
is allowed. This statement applies to quasi-steady systems and
may break down for galaxies that accumulate their gas mass reservoir via
a series of violent mergers or if an external perturbation drives the system 
out of equilibrium and triggers star formation in some parts of the galaxy,
as discussed later in Section~\ref{steadystate}.

\begin{figure*}
 \centering
  \includegraphics[width=15cm]{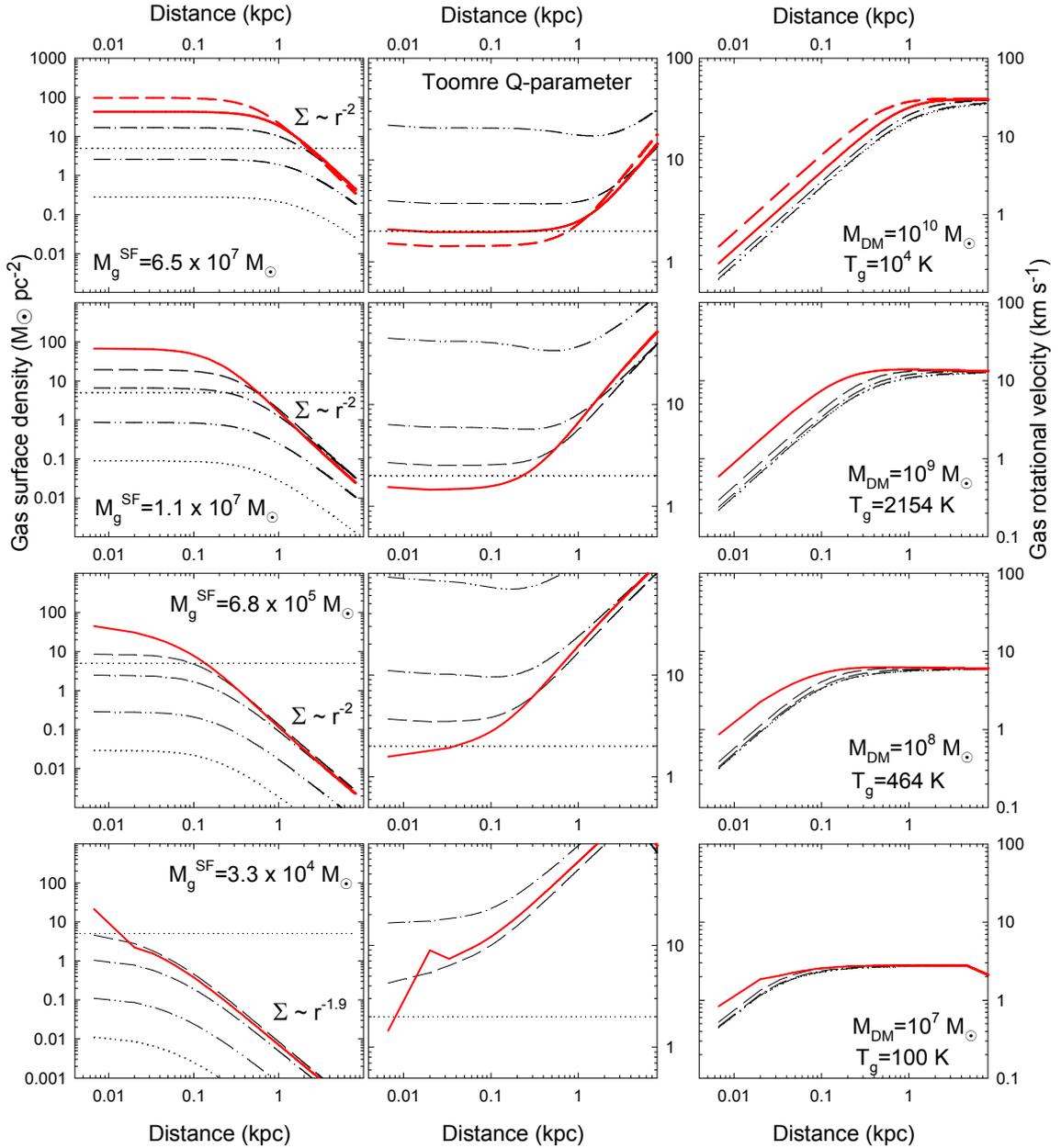}
      \caption{Same as Fig.~\ref{fig2} only for the gas temperature 
$T_{\rm g}\propto M_{\rm DM}^{2/3}$,
      with the actual values of $T_{\rm g}$ indicated in the 
right column.}
\label{fig5}
\end{figure*}

An alternative solution is that galaxies can 
cool to sufficiently low temperatures to warrant a more compact and dense gas configuration.
Figure~\ref{fig5} shows gas surface densities $\Sigma$ (left column),
Toomre $Q$ parameters (middle column), and gas rotation velocities
$v_\phi$ (right column) for the same parameters as in
Figure~\ref{fig2} but for $T_{\rm g} \propto M_{\rm DM}^{2/3}$ 
as described by eq.~(\ref{scaling}), with
the corresponding gas temperatures indicated for every $M_{\rm DM}$ in
the right column.  One can see that the gas surface density profiles
are considerably steeper for cooler gas disks 
and are characterized by approximately the same power law $r^{-2}$ in the outer
regions. Furthermore, the transition radius from a
near-constant surface density to the sloped one decreases with 
mass\footnote{Diminishing pressure support against gravity in models with smaller $T_{\rm g}$ 
is compensated by an increase in the gas density slope.}, 
in contrast to models with the DM-mass-independent gas temperature $T_{\rm g}$ (Fig.~2).

The star formation criteria~(\ref{Toomre}) and (\ref{KS}) are satisfied only in the inner
parts of our model galaxies, with the size of the star formation region
shrinking to a few tens of parsecs for models with $M_{\rm
DM}\le10^8~M_\odot$.  Moreover, the value of $M_{\rm g}^{\rm SF}$ notably decreases for lower
mass DM halos. These very compact starburst regions are not unusual in low mass star forming galaxies.
For instance, the galaxy SBS 0335-052 has a star forming radius of
only $\sim$~20~pc \citep[see e.g.][]{Takeuchi03}.

As the blue dashed line in the top panel of Fig.~\ref{fig4}
demonstrates, models with $T_{\rm g}\propto M_{\rm DM}^{2/3}$ 
have much smaller $M_{\rm g}^{\rm min}$ 
than models with $T_{\rm g}=10^4$~K.  Moreover, models with
$T_{\rm g}\propto M_{\rm DM}^{2/3}$ are characterized by $M_{\rm
  g}^{\rm min}<M_{\rm b}$, meaning that such galaxies may have enough
gas reservoir to achieve a state in which SF is allowed.  
However, the required gas temperatures are quite low, 
especially for $M_{\rm DM} \le 10^{8}~M_\odot$.
Moreover, the obtained rotation curves (third column in Figure \ref{fig5}) are much flatter 
for models with $M_{\rm DM} \le 10^{8}~M_\odot$ 
than for their more massive counterparts. 
We conclude that models with   $T_{\rm g}\propto M_{\rm DM}^{2/3}$ can provide
a better agreement with the $\Lambda$CDM predictions but at the cost of 
a worsening agreement with observations for the low DM mass models.

\subsection{The effect of varying rotation support}
\label{spin}
Decreasing the amount of rotation leads to more compact and dense gas
configurations as the pressure forces start to play an ever increasing
role in the support against gravity.  Therefore, one can expect that
lowering the spin parameter would produce steady-state
gaseous configurations with steeper gas surface density profiles (provided that
the gas temperature is constant). 

\begin{figure}
  \resizebox{\hsize}{!}{\includegraphics{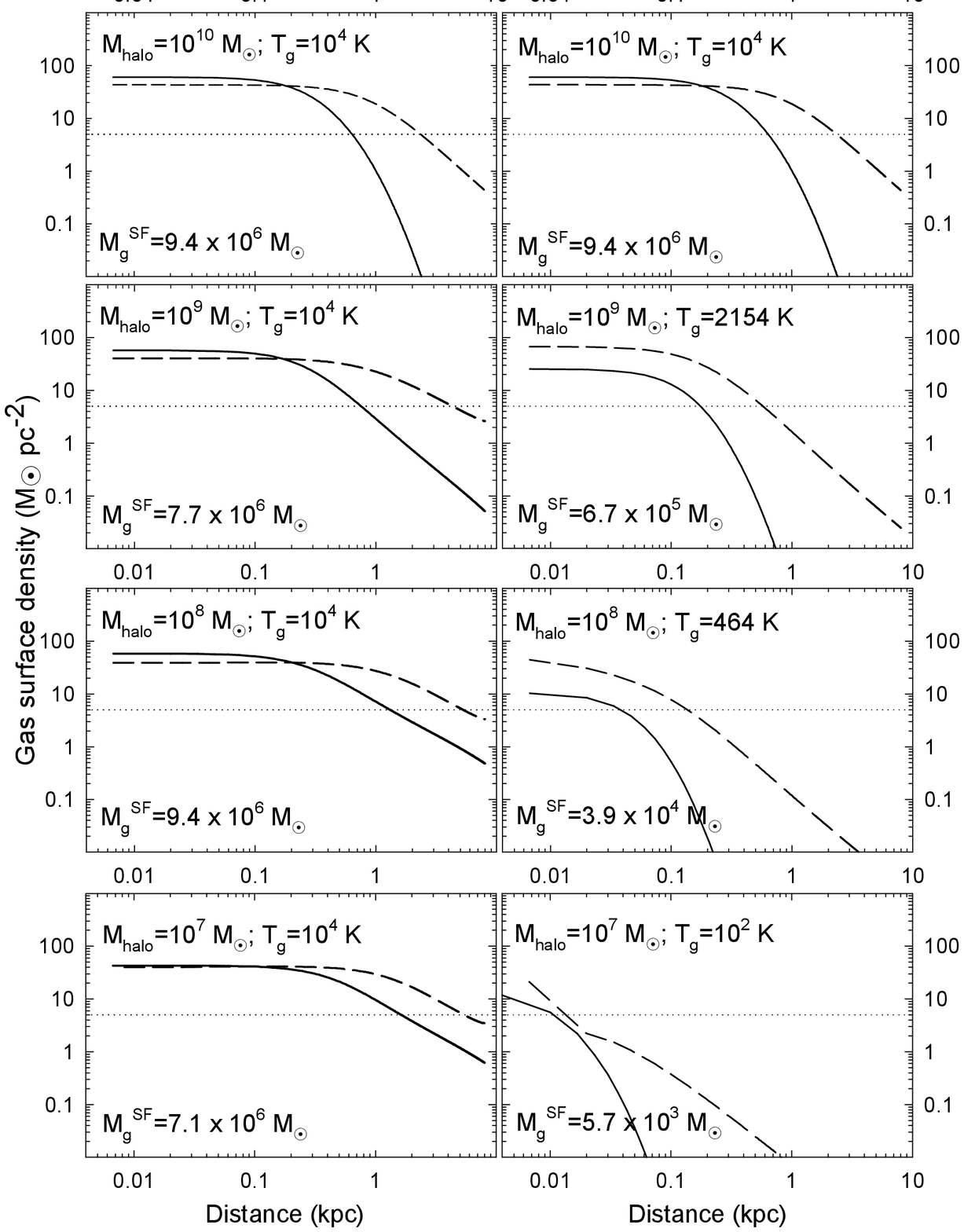}}
      \caption{Comparison of the gas surface density profiles for models with 
      the spin parameter $\alpha=0.5$ (solid lines) and $\alpha=0.9$ (dashed lines).
      The DM halo mass $M_{\rm DM}$ and the gas temperature $T_{\rm g}$  are indicated in 
      every panel. The gas mass $M_{\rm g}^{\rm SF}$ with number density greater than 
      1.0~cm$^{-3}$ is given only for the $\alpha=0.5$ models.}
         \label{fig5a}
\end{figure}

This effect is illustrated in Figure~\ref{fig5a} showing 
the gas surface density distributions for models with $\alpha=0.5$ (solid lines) and $\alpha=0.9$
(dashed lines). 
For the sake of convenience, we compare only the star-formation-allowed models\footnote{Strictly
speaking, the $M_{\rm DM}=10^7~M_\odot$ and $T_{\rm g}=10^2$~K case has no star-formation-allowed 
models since the third criterion, i.e., $M_{g}^{\rm SF}>10^4~M_{\odot}$ is not 
fulfilled even for the highest $n_{0,0}$ model. We relax this requirement in this particular case 
since the other two SF criteria are nevertheless satisfied. }
with the smallest value of $M_{\rm g}^{\rm min}$. In the case of $\alpha=0.9$, such 
models are distinguished by red solid lines in Figures~\ref{fig2} and \ref{fig5}.
As one can see, models with smaller $\alpha$ are characterized by more compact gas disks with
steeper density profiles at large radii. This effect takes place 
because the $\alpha=0.5$ models compensate for a smaller rotation support with 
steeper gas density (and pressure) gradients. 
As a result, these models also have smaller $M_{\rm g}^{\rm SF}$ than the 
corresponding $\alpha=0.9$ models\footnote{We note
that some of the $\alpha=0.5$ models have higher gas densities in the innermost regions,
which is the result of a rather coarse grid of models considered in the paper. Nevertheless,
the size of this region is rather small ($\sim 0.2$~kpc) and most of the gas mass is still 
concentrated at large radii. }.

Figure~\ref{fig4} illustrates the effect of a smaller rotation support
against gravity. The thin red lines show the data for $\alpha=0.5$, with
other parameters being identical to those of the reference model. It is evident that the
minimum gas mass for star formation $M_{\rm g}^{\rm min}$ is
substantially lower in galaxies with smaller values of $\alpha$. The
effect is particularly strong for galaxies with $T_{\rm g}\propto
M_{\rm DM}^{2/3}$. Lowering further the spin parameter to
$\alpha=0.25$ yields only a factor of two decrease in $M_{\rm g}^{\rm
  min}$ at most.  We point out that the minimum gas mass for star
formation $M_{\rm g}^{\rm min}$ remains still much greater than the
available baryonic mass $M_{\rm b}$ for galaxies with $M_{\rm DM}\le
10^8~M_\odot$ and $T_{\rm g}=10^4$~K.  

\subsection{Variations in the DM halo form}
\label{NFWhalo} 
As already mentioned, the mass and shape
  of DM halos in DGs are still very uncertain and the
  problem is highly debated in the literature.  If the DM density at 
  small galactocentric radii $\varpi$ is approximated by a power law 
($\rho_{DM}\propto \varpi^{-\beta}$), a value of $\beta$ 
close to or slightly smaller   than one (i.e., cusps) is obtained 
in numerical simulations based on the $\Lambda$CDM theory 
\citep{NFW,nava10}. On the other hand, observations suggest
  values of $\beta$ close to zero, implying the presence of cores with near-constant 
  DM density at small $\varpi$ \citep{debl01,spek05}.  This unsolved mismatch between observations
  and models is commonly referred to as the ``cusp-core problem''.

\begin{figure}
  \resizebox{\hsize}{!}{\includegraphics{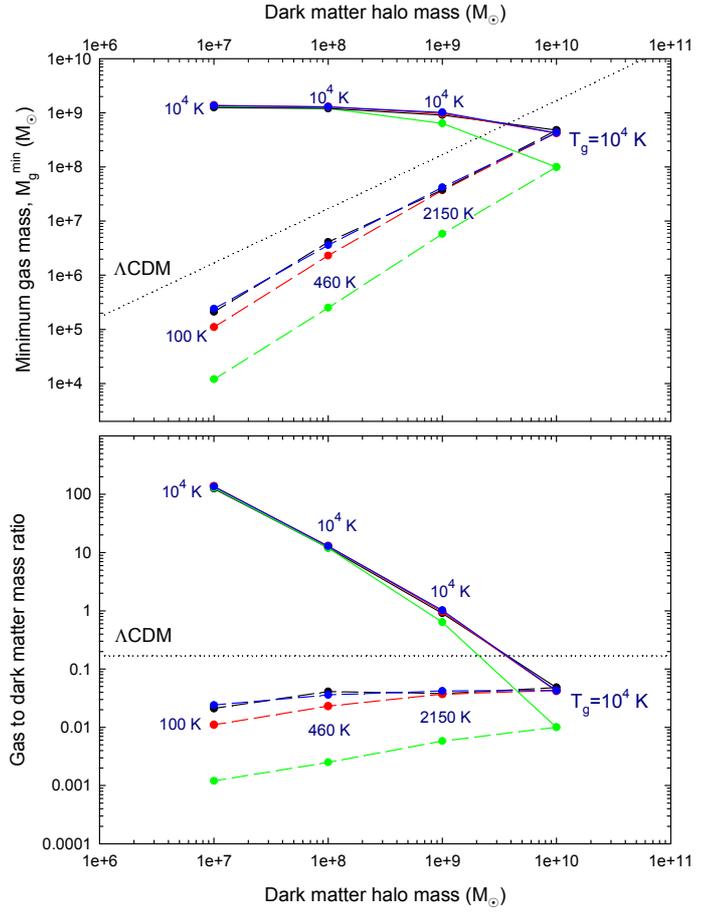}}
      \caption{{\bf Top.} Minimum gas mass for star formation $M_{\rm g}^{\rm min}$
      versus DM halo mass $M_{\rm DM}$ for the reference model (blue lines), 
      models with a cuspy DM halo (red lines), models with pre-existing stellar 
      disk (black lines), and models at redshift $z_{\rm gf}=2.0$ (green lines). The solid lines show
      models with the gas temperature $T_{\rm g}=10^4$~K, whereas dashed lines do that for 
      models with $T_{\rm g}\propto M_{\rm DM}^{2/3}$ (see equation~\ref{scaling}). 
      Numbers indicate the corresponding gas temperatures for every model.
      The dotted line plots the cosmological relation between the baryonic and DM mass
      as suggested by the $\Lambda$CDM theory. {\bf Bottom.} The corresponding 
      $M_{\rm g}^{\rm min}/M_{\rm DM}$ versus
      $M_{\rm DM}$ relations. }
         \label{fig6}
\end{figure}

Our reference quasi-isothermal DM halos are cored DM profiles.
In this section we estimate the effect that a cuspy DM halo advocated
by \citet{NFW} and described by
equations~(\ref{cuspyhalo})-(\ref{cparameter}) may have on the values
of $M_{\rm g}^{\rm min}$. Figure~\ref{fig6} compares the minimum gas
masses $M_{\rm g}^{\rm min}$ obtained in the reference model for the quasi-isothermal DM halo
(blue lines) with those calculated for the NFW halo (red lines). As
one can see, the difference is negligible for models with $T_{\rm
g}=10^4$~K and is minimal for models with $T_{\rm g}\propto M_{\rm
DM}^{2/3}$.  Therefore, the form of the DM halo is not expected to play 
a significant role in determining the minimum gas mass for star formation  
as far as the total mass of the DM halo is the same.

\subsection{The effect of pre-existing stellar disk}
\label{stars}
So far we have considered model galaxies that consist of a gaseous disk and 
DM halo. However, real star-forming galaxies in the Local Universe almost 
always have a pre-existing stellar disk, which
may affect the form of the gaseous disk via the stellar gravitational
potential.  To explore the extent of this effect, we assume that our galaxies in the 
reference model have a burst of star formation that turns 10\% of the gas content into stars. 
Star formation takes place in parts of the gas disk that obey the three criteria
for star formation $\Sigma>\Sigma_{\rm c}=5~M_\odot$~pc$^{-2}$ and $Q_{\rm T}<Q_{\rm cr}=2.0$.
We then construct a new equilibrium gas disk in the combined gravitational potential
of gas, stars, and DM halo. The spin parameter of the stellar disk is
assumed to be equal to that of the gaseous disk. This assumption is justified
if the gas sound speed is comparable to the stellar velocity dispersion, which is often true
for DGs.

The black lines in Figure~\ref{fig6} show the minimum gas mass $M_{\rm
  g}^{\rm min}$ for new models that take into account a recent burst of
  star formation.  It is evident that the stellar disk has
little effect on the value of $M_{\rm g}^{\rm min}$ 
irrespective of the DM halo mass (the black line is almost 
indistinguishable from the blue line showing the reference model). 
This is explained by the fact that 
the mass of the stellar disk $M_{\rm s}$ is only a small addition
to the total mass budget. For instance, in models with $M_{\rm DM}=10^{10}~M_\odot$,
$M_{\rm s}=0.07~M_{\rm g}^{\rm min}$ and $M_{\rm s}=0.0028~M_{\rm DM}$. In models
with $M_{\rm DM}=10^7~M_\odot$, $M_{\rm s}=0.047~M_{\rm g}^{\rm min}$ and 
$M_{\rm s}=4.1~M_{\rm DM}$ (in both cases, $T_{\rm g}=10^4$~K). 
Unless many repetitive bursts of star formation take place with the integrated
star formation efficiency considerably exceeding 0.1 (leading to a 
significant increase in the star to gas mass ratio), we do not expect the gravitational
potential of the stellar disk to affect our results. However,
we should note that the stellar feedback may drive the 
gas disk out of equilibrium, thus affecting our estimates of $M_{\rm g}^{\rm min}$.

\subsection{Galaxies at higher redshifts}
\label{highzgf}
In this section, we study the dependence of $M_{\rm g}^{\rm min}$ on
the redshift of galaxy formation $z_{\rm gf}$.  DM halos of the same mass
at larger redshifts are more compact and one may expect that this
could affect the shape of the gas disk and hence the value of $M_{\rm
  g}^{\rm min}$. We modify
equations~(\ref{scalelength})-(\ref{virial}) to include the dependence
on $z_{\rm gf}$ \cite[e.g.][]{Fujita03}
\begin{eqnarray}
\label{scalelengthz0}
r_{\rm 0}(z_{\rm gf})&=& r_0  \left({\Omega_0 \over \Omega(z_{\rm gf})}\right)^{-1/3} (1+z_{\rm gf})^{-1}
\,\, \mathrm{kpc}, \\
\rho_{\rm 0}(z_{\rm gf})&=& \rho_0 \left({\Omega_0 \over \Omega(z_{\rm gf})}\right) (1+z_{\rm gf})^{3}  \,\, M_\odot \, \mathrm{kpc}^{-3}, \\
\label{virialz0}
\varpi_{\rm vir}(z_{\rm gf})&=&\varpi_{\rm vir} \left({\Omega_0 \over \Omega(z_{\rm gf})}\right)^{-1/3} (1+z_{\rm gf})^{-1} \,\, \mathrm{kpc},
\end{eqnarray}
where $\Omega_0\equiv\Omega(z_{\rm gf}=0)$ and $\Omega(z_{\rm gf})$ is defined as
\begin{equation}
\Omega(z_{\rm gf})= {\Omega_m (1+z_{\rm gf})^3 \over \Omega_m (1+z_{\rm gf})^3 + \Omega_\Lambda},
\end{equation}
with $\Omega_m$ and $\Omega_\Lambda$ set to 0.24 and 0.76, respectively.

The green lines in the upper panel of Figure~\ref{fig6} show the
minimum gas mass for star formation $M_{\rm g}^{\rm min}$ as a
function of $M_{\rm DM}$ in the reference model but with $z_{\rm gf}=2.0$.  
In compliance with the downsizing, low-mass galaxies have a median redshift of star
 formation smaller than large objects and a value of $z_{\rm gf}=2$ well
  represents an average galaxy formation redshift for DGs
  \citep[see e.g.][]{Thomas05,Cattaneo11}.
We also note that $z_{\rm gf}=2.0$ yields roughly a factor of two more
compact and a factor of seven denser DM halos. 

As one can see, the effect of a higher redshift is quite pronounced for models with
$T_{\rm g}\propto M_{\rm DM}^{2/3}$ 
(shown by the green dashed lines), producing almost a factor of 10
lower values of $M_{\rm g}^{\rm min}$ than in the reference model (blue dashed lines). 
However, for models with
$T_{\rm g}=10^4$~K, the resulting values of $M_{\rm g}^{\rm min}$
quickly approach those of the $z_{\rm gf}=0$ models (blue lines) for low values
of $M_{\rm DM}$.

\begin{figure}
  \resizebox{\hsize}{!}{\includegraphics{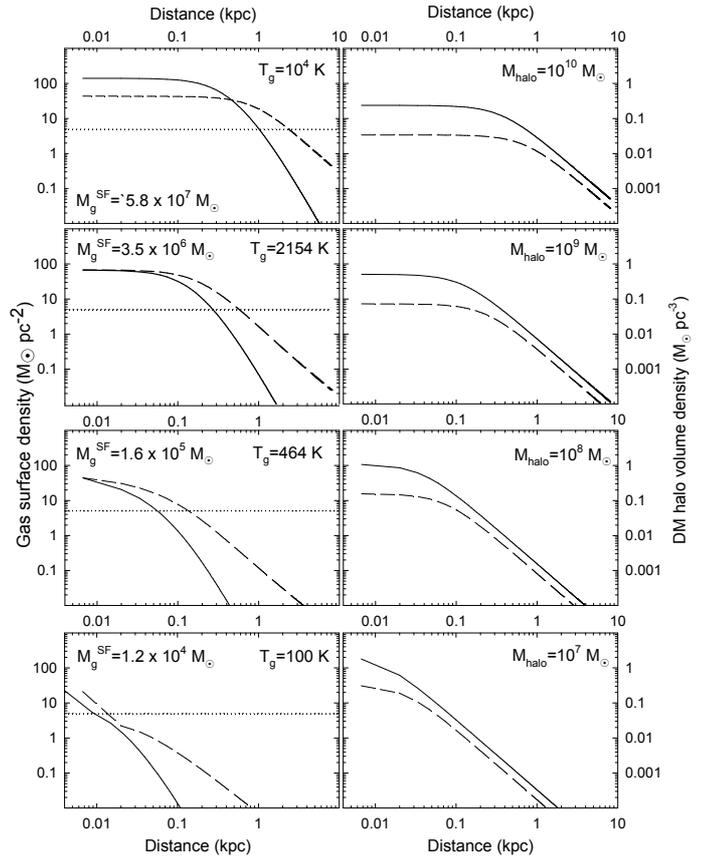}}
      \caption{{\bf Left column}. Comparison of the gas surface density in the reference model (dashed
      lines) with that of the $z_{\rm gf}=2.0$ model (solid lines). The dotted line marks the 
      critical surface density for star formation $\Sigma_{\rm c}=5~M_\odot$~pc$^{-2}$. 
      Gas temperature is indicated in every panel and the DM halo mass is shown in the 
      corresponding panels on the right. The gas mass $M_{\rm g}^{\rm SF}$ 
      with number density greater than 1.0~cm$^{-3}$ is given only for the $z_{\rm gf}=2.0$ models.
      {\bf Right column.} Volume density profiles of the quasi-isothermal DM halos of various mass (as
      indicated in each panel) at $z_{\rm gf}=0$ (dashed lines) and $z_{\rm gf}=2.0$ (solid lines). }
         \label{fig6b}
\end{figure}

The lowering of $M_{\rm g}^{\rm min}$ can be understood if we compare
the radial density profiles of the DM halos of equal mass at different redshifts. 
The right panel in Figure~\ref{fig6b} presents the volume density
 $\rho_{\rm qis}$ of the quasi-isothermal DM halo as a function of radial distance 
 $\varpi$ for the reference model 
 ($z_{\rm gf}=0$, dashed lines) and the $z_{\rm gf}=2.0$ model (solid lines).
The corresponding masses of the DM halos are indicated in each panel.
One can see that the DM halos at $z_{\rm gf}=2.0$ are characterized by higher
densities than their $z_{\rm gf}=0$ counterparts. 
This causes gaseous disks at higher redshifts to assume steeper density profiles to  
compensate an increasing gravity force of the DM halo. This effect is illustrated
in the left panel of Figure~\ref{fig6b}, which compares the gas surface density
profiles in the star-formation-allowed models at $z_{\rm gf}=0$ (dashed lines) 
and the $z_{\rm gf}=2.0$ model (solid lines). All other model parameters are those of the reference
model and the gas temperature is indicated in each panel. Indeed, the $z_{\rm gf}=2.0$ models
are characterized by significantly more compact gaseous disks than their $z_{\rm gf}=0$ counterparts,
which results in systematically lower values of $M_{\rm g}^{\rm min}$.

\section{Reconciling the gas to DM ratio with the $\Lambda$CDM predictions}
\label{LCDM}
\begin{figure}
  \resizebox{\hsize}{!}{\includegraphics{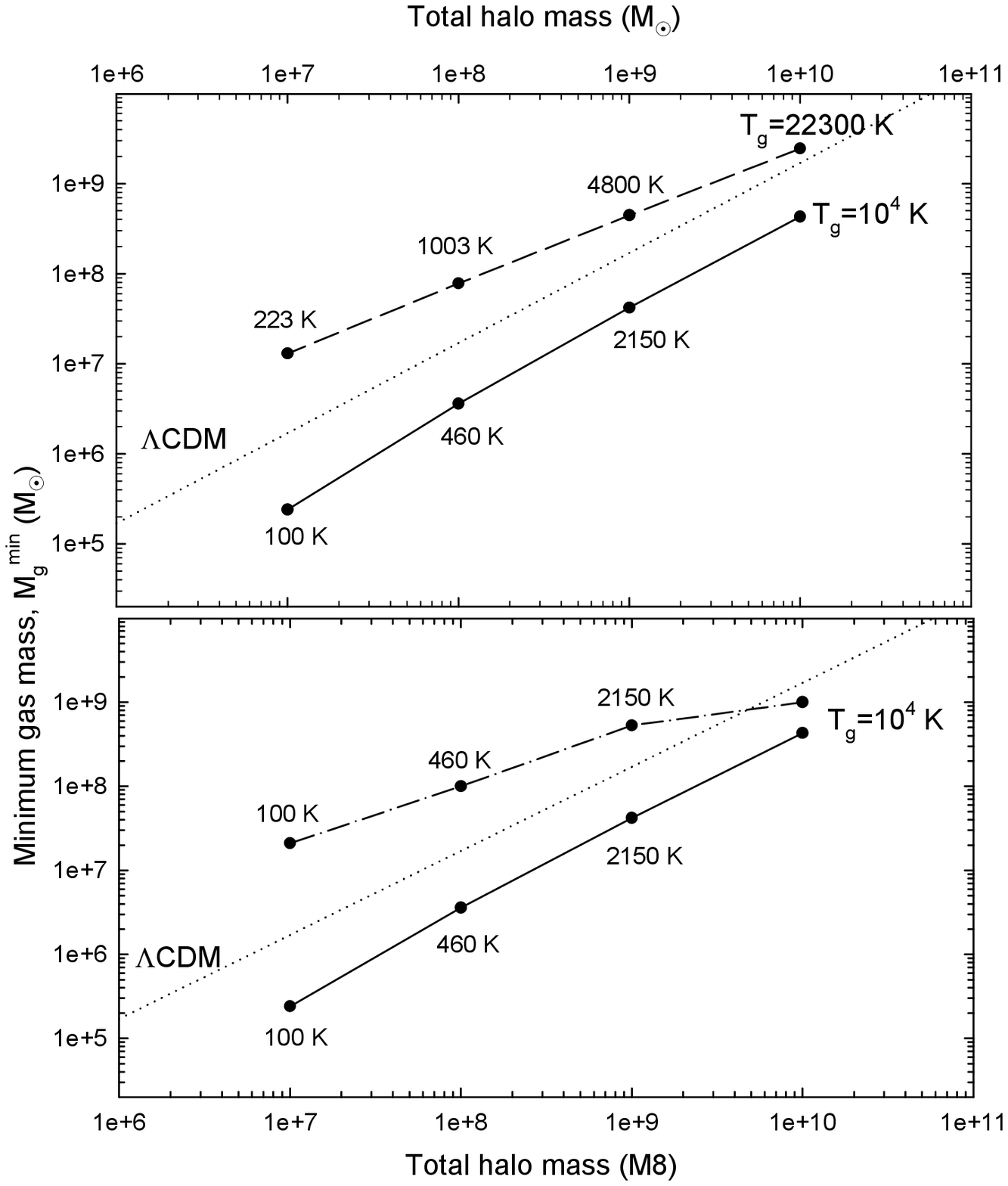}}
      \caption{{\bf Top.} Effect of changing the zero-point of the $T_{\rm g}\propto M_{\rm DM}^{2/3}$
      scaling law. The solid lines present the $M_{\rm g}^{\rm min}$--$M_{\rm DM}$ relation for the
      reference model with scaling described by equation~(\ref{scaling}), whereas the 
      dashed line does that for the scaling
      given by equation~(\ref{scaling2}). The corresponding gas temperatures are indicated for each
      model. {\bf Bottom.} The effect of non-thermal gas support against gravity. Solid lines show
      the $M_{\rm g}^{\rm min}$--$M_{\rm DM}$ relation for the reference model (zero non-thermal support),
      whereas the dashed line does that for the models with the effective gas pressure 
      three times greater than the gas kinetic pressure. The dotted line plots the 
      cosmological relation between the baryonic and DM mass
      as suggested by the $\Lambda$CDM theory.  }
         \label{fig7}
\end{figure}

Figures~\ref{fig4} and \ref{fig6} demonstrate that the minimum gas mass for star formation 
$M_{\rm g}^{\rm min}$
may greatly exceed the DM halo mass (and, of course, the gas to DM mass ratio can be largely above the
suggested cosmological value of 0.17) for models with $M_{\rm DM}\le10^{9}~M_\odot$ and 
$T_{\rm g}=10^4$~K. This apparent contradiction between the 
baryon-to-DM ratios deduced by our models and the cosmologically inferred one
should not be a major source of concern at the present stage.
In fact, we know already that some objects in the Universe do have
baryon-to-DM ratios larger than 0.17 (e.g., globular clusters, high velocity
clouds, and probably also dEs and dIrrs, see the Introduction).
At the same time, Figures~\ref{fig4} and \ref{fig6} show that  models in which 
$T_g \propto M_{DM}^{2/3}$ (as suggested by the virial relations) neatly reproduce
a constant gas to DM mass ratio, although below the cosmological value
of 0.17. 

In this section, we explore whether the cosmological value 
can be reproduced by our models with either a
different choice of the zero point of the $T_{\rm g}$--$M_{\rm DM}$ relation
or by introducing some non-thermal support against gravity in equation~(\ref{equilib}). 
To explore the first possibility, we 
choose the following scaling law between the gas temperature and the DM halo mass
\begin{equation}
\label{scaling2}
T_{\rm g}=4.8\times10^{-3} M_{\rm DM}^{2/3},
\end{equation}
which yields roughly a factor of two higher gas temperatures than
equation~(\ref{scaling}), i.e., $T_{\rm g}=2.23\times10^4$~K 
for $M_{\rm DM}=10^{10}~M_\odot$,   $T_{\rm g}=0.48\times10^4$~K for $M_{\rm DM}=10^{9}~M_\odot$,
$T_{\rm g}=1.03\times10^3$~K for $M_{\rm DM}=10^{8}~M_\odot$, and $T_{\rm g}=223$~K for 
$M_{\rm DM}=10^{7}~M_\odot$.

The top panel in Figure~\ref{fig7} shows the minimum gas mass for star formation 
$M_{\rm g}^{\rm min}$
as a function of the DM halo mass $M_{\rm DM}$. The dashed line presents the data for the new scaling
law described by equation~(\ref{scaling2}), while the 
solid line does that for the old scaling given by equation~(\ref{scaling}). As one can see, 
the new scaling law yields somewhat higher gas masses than a cosmological value and
the slope of the model  $M_{\rm g}^{\rm min}$--$M_{\rm DM}$ 
relation is somewhat shallower than the cosmological one (dotted line).
On the other hand, the old scaling law yields $M_{\rm g}^{\rm min}$ that are somewhat smaller
than the cosmological values.
This means that varying the zero point of the $T_{\rm DM}\propto M_{\rm DM}^{2/3}$ relation, one 
can in principle achieve a good agreement with the $\Lambda$CDM theory. 

To explore the second possibility, we assume that our model gas disk has a non-thermal
support against gravity in the form of turbulence and magnetic pressure. For the sake of simplicity,
we assume equipartition between the gas kinetic pressure $P$ and the 
two sources of non-thermal support (but see \citet{Cox05}), which yields the 
effective gas pressure $P_{\rm eff}=3 P=3 \rho_{\rm g} \sigma_{\rm g}^2$
(which would correspond to an effective velocity dispersion 
$\sigma_{\rm eff}=\sqrt{P_{\rm eff}/\rho_{\rm g}} =\sqrt{3}\sigma_{\rm g}$). 
The dash-dotted line in the bottom panel
of Figure~\ref{fig7} presents the $M_{\rm g}^{\rm min}$--$M_{\rm DM}$ relation for the case with
the non-thermal support. It is evident that the corresponding gas masses increase considerably, in particular
for models with $M_{\rm DM}\le 10^9~M_\odot$. The matter is that introducing the non-thermal support
we effectively increase the gas pressure and the corresponding gas surface density profiles become
shallower as compared to those without non-thermal support. This results in an overall increase
in the gas mass of a steady-state gaseous configuration. Figure~\ref{fig7} also suggests that
varying the magnitude of the non-thermal support one can achieve a good agreement with 
the cosmological mass of baryons given by the dotted line, particularly for galaxies
with $M_{\rm DM}\le10^{10}~M_\odot$.

\section{Implications for the evolution of dwarf galaxies}
\label{discuss}
In this work, we have presented numerical solutions for equilibrium
configurations of model galaxies made up of gas, stars and a DM halo in
the combined gravitational potential of each of these components.  
The properties of these equilibrium configurations and, in particular, the 
minimum gas mass needed to achieve a state with allowed star formation $M_{\rm g}^{\rm min}$,
have been considered in detail.

In future more detailed works, we will be using our derived equilibrium configurations 
as initial setups of galaxy models for which we will numerically study
the detailed chemical and dynamical evolution, and also the effect of SF feedback.
The interest in simulating the evolution of DGs is steadily growing in the
last years.  The reason is that $\Lambda$CDM theories of structure
formation predict that dwarf galaxy-sized objects are the first
virialized structures in the Universe.  Moreover, the study of star
formation and feedback in DGs is in many respects much
simpler than in large spiral galaxies.  Although studies of DGs
 in a cosmological context are more numerous and detailed than
in the past \citep[e.g.][]{kaza11,sales11,pilk11,gove10}, still they
do not have enough spatial resolution to analyze in detail the
internal evolutionary processes.  A lot in resolution can be gained by
zooming in and re-simulating small chunks of a large cosmological box
\citep{mart09,sawa11}, but still the best way to accurately simulate a
dwarf galaxy is by numerically studying it as a single isolated entity 
\citep{schr11,sb10,revaz09} although in reality they are subject 
to various environmental effects.

Although a quantitative comparison between our
predictions and observations in DGs requires taking into account the feedback from 
star formation, yet our modeling can give us some insight 
as to the expected evolution of DGs. First, we note that for some models
(especially the low-mass ones with T=10$^4$ K, see Figure \ref{fig2},
bottom row of panels) star formation is not expected to occur in the
center of the galaxy, but only in a shell with inner radius between
$\simeq$ 100 pc and $\simeq$ 1.0~kpc. 

\begin{figure}
  \resizebox{\hsize}{!}{\includegraphics{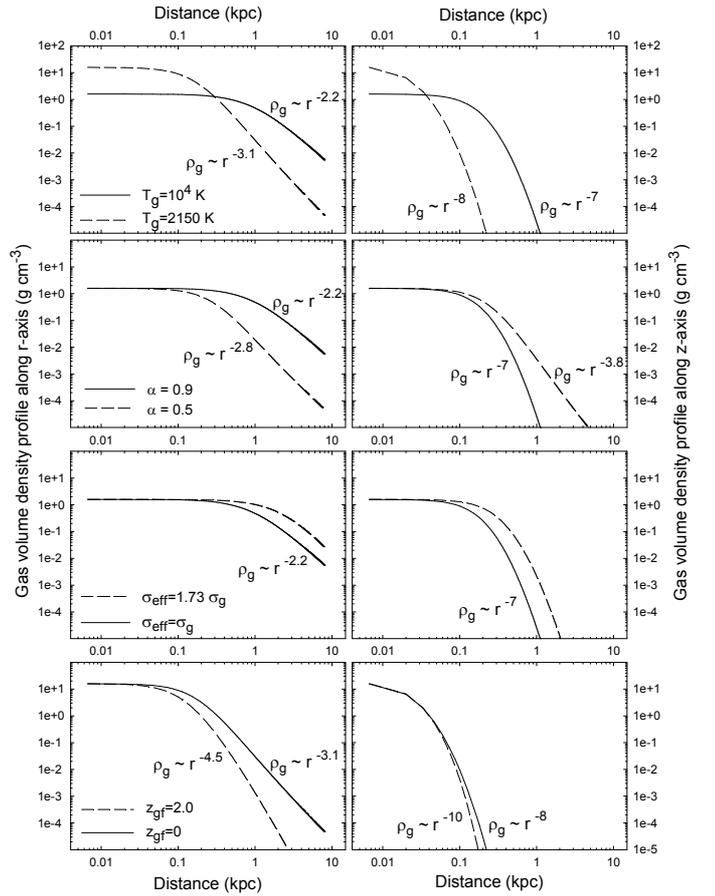}}
      \caption{Comparison of the gas volume density profiles along the
      $r$-axis (left column) and $z$-axis (right column). In particular,
      models in the top row are characterized by different gas temperatures $T_{\rm g}$ 
      (as indicated), the second top row by different spin parameters $\alpha$, the third
       top row by different effective gas dispersions $\sigma_{\rm eff}$, and the bottom row
       by different redshift for galaxy formation $z_{\rm gf}$. All other parameters for every model
       are the same (see text for more details).}
         \label{fig8}
\end{figure}

Second, the evolution of SN-driven shells is known to depend on the
gas density distribution, which in turn is sensitive to the initial
conditions in a dwarf galaxy. For instance, the Rayleigh-Taylor 
instability in the shell grows faster for steeper gas density profiles.
Figure~\ref{fig8} compares the gas {\it volume}
density distributions $\rho_{\rm g}$ in star-formation-allowed models with different $T_{\rm g}$, 
$\alpha$, $\sigma_{\rm eff}$, and $z_{\rm gf}$. In particular, the left/right columns 
show the radial profiles of $\rho_{\rm g}$ taken along the $r$-/$z$-axes.
For the sake of conciseness, we consider only models with $M_{\rm DM}=10^9~M_\odot$,
models with other DM halo masses show a similar behaviour. Models in the top row
are characterized by $\alpha=0.9$, $\sigma_{\rm eff}=0$, and $z_{\rm gf}=0$ (but different 
$T_{\rm g}$ as indicated in the Figure), models in the second top row by $T_{\rm g}=10^4$~K,
$\sigma_{\rm eff}=0$, and $z_{\rm gf}=0$ (but different $\alpha$), models in the third 
row by $T_{\rm g}=10^4$~K, $\alpha=0.9$, and $z_{\rm gf}=0$ (but different $\sigma_{\rm eff}$),
and models in the bottom row by $T_{\rm g}=2150$~K, $\alpha=0.9$, $\sigma_{\rm eff}=0$ (but  different
$z_{\rm gf}$). 

It is evident that taking a smaller gas temperature $T_{\rm g}$ or spin parameter results
in equilibrium gas disks with a steeper tail of the gas volume density distribution.
A similar effect is observed for models with a higher redshift for galaxy formation 
$z_{\rm gf}$. On the other hand,
models with and without non-thermal support have similar radial profiles of $\rho_{\rm g}$.
For every model considered, the vertical profiles of $\rho_{\rm g}$ are steeper than those taken along
the horizontal axis, suggesting a blow-out effect along the rotational axis.
The variety of possible model realizations implies that the evolution of DGs
after the onset of star formation may follow different pathways depending on the initial
conditions in the gas disk, even for the same DM halo mass.

\section{Model caveats}
\label{caveats}

\subsection{Steady-state gaseous disks}
\label{steadystate}
A steady-state model is a first-order approximation to DGs galaxies.
Various effect such as stellar feedbacks, non-axysimmetric density waves, and, in particular, 
external perturbations may drive DGs out of equilibrium.
These phenomena can trigger star formation in otherwise quiescent gas disks and affect our 
derived values of $M_{\rm g}^{\rm min}$. In order to estimate the possible magnitude 
of such effects, we focus on perturbations with the conservation of the total gas 
mass\footnote{Perturbations without conservation of the total mass, such as mergers or ram pressure
stripping, can obviously affect our conclusions by changing the total available 
gas mass budget.} and refer to star-formation-inactive models plotted in Figure~\ref{fig2} by black
lines. 
It is evident that positive perturbations in $\Sigma$ by a factor of 5--100 are needed to drive 
these models to the star formation threshold. 

If perturbations of such amplitude are possible, then the critical gas mass required for star formation may be significantly lower.
Indeed, as Figure~\ref{fig3} demonstrates, the total gas mass $M_{\rm g}$ of an equilibrium 
configuration declines with decreasing $n_{0,0}$. The filled squares mark the star-formation-allowed
models, while the open circles correspond to the star-formation-inactive
ones. If the $n_{0,0}=0.01$cm$^{-3}$ 
models can be pushed to the star formation threshold, then the minimum mass for star formation 
$M_{\rm g}^{\rm min}$ may be as low as $10^7~M_\odot$ for the $M_{\rm DM}=10^{10}~M_\odot$ model. 
This corresponds to almost a factor of 40 decrease
in the value of $M_{\rm g}^{\rm min}$ as compared to the star-formation-allowed 
model with $n_{0,0}=5$~cm$^{-3}$. 
We note, however, this effect becomes considerably less pronounced
for models with smaller DM halo masses. For instance, the corresponding decrease
in $M_{\rm g}^{\rm min}$ for the $M_{\rm DM}=10^7~M_\odot$ model is only a factor of 2.5.   
We conclude that high-amplitude density perturbations of the equilibrium state can 
significantly affect our estimates of $M_{\rm g}^{\rm min}$ only for models with 
$M_{\rm DM}\ga 10^9~M_\odot$ but are of rather low significance for models with 
$M_{\rm DM}\la 10^8~M_\odot$.

\subsection{Multi-phase interstellar medium}
In this study we have neglected the fact that the interstellar medium
consists of various phases with usually different temperatures and
considered a single-phase medium with some typical temperature $T_{\rm g}$. 
Although cores of molecular clouds 
(where star formation occurs) have temperatures much lower than $10^4$~K, 
the latter value must be seen as a mass-weighted mean temperature
within each computational cell (which has a size much larger than the
cores of molecular clouds). Indeed, the hot ($\sim 10^6$~K) and 
cold ($\sim \mathrm{a~few} \times 10$~K)
gas phases usually amount to about 1\% and 10\%
of the total gas mass reservoir, respectively 
(for the SF efficiency of $\sim 10\%$), meaning that the mean temperature
is mostly determined by the warm gas phase. 
In this sense, $T_{\rm g}=10^4$~K represents a sort of
an upper limit because it is impossible that a computational cell hosting
(star-forming) regions with temperatures of few tens of K can have an
average temperature significantly above 10$^4$~K. We note that lower than $10^4$~K 
mean temperatures are, of course, possible, provided efficient cooling and low SF feedback.

One may argue that even though some models for $T_{\rm g}=10^4$~K 
may have difficulty to achieve critical densities for
star formation, the differentiation into a multi-phase medium may
eventually push a local region to star formation and that may trigger
a chain reaction through the bulk of a galaxy. To account for this
possibility, we have adopted an empirical star formation threshold by
\cite{Kennicutt08}. While the Toomre $Q$ parameter criterion is based
on the gravitational properties of a single-phase medium, the
Kennicutt's criterion is based on observations of real multi-phase
galaxies and hence takes implicitly into account the possibility of
phase differentiation discussed above.
In a comprehensive review, \citet{Hensler08} has discussed the advantages 
of a multi-phase treatment
of the ISM in star-forming galaxy disks and emphasized the limitations of
single gas-phase description.

\subsection{Star formation criteria}
In this paper, we have adopted three SF criteria, which are based on theoretical
arguments, i.e., the Toomre gravitational stability criterion~(\ref{Toomre}), and empirical evidence,
i.e., the Kennicutt-Schmidt law~(\ref{KS}). These SF criteria are not without limitations and
observations suggest that SF can occur even in the Toomre-stable regions with $Q_{\rm T}\ga Q_{\rm c}$,
e.g., near the galactic center where the gravitational stability may be determined by the rate of sheer
rather than by the magnitude of epicycle motions \citep{Vor03}. Moreover, a few galaxies in the Kennicutt's
(2008) sample harbour SF below the adopted gas density threshold of $\Sigma_{\rm c}=5~M_\odot$~pc$^{-2}$.
At the lower end of the KS correlation the scatter of measurements also widens,
because the SF rate fluctuates more stochastically and e.g. starbursting DGs are systematically
located above the relation \citep[see e.g.][]{Hensler12}. In addition, it is
well documented that the KS relation is tighter when the SF rate is correlated with the
molecular hydrogen H$_2$ \citep{Kennicutt07,Bigiel}.

Nevertheless, we can use the same arguments as in Section~\ref{steadystate} to show 
that a factor of 
ten variation in the adopted value of $\Sigma_{\rm c}$ can significantly affect our estimates of
$M_{\rm g}^{\rm min}$ only in models with the DM halo mass $M_{\rm DM}\ga 10^9~M_\odot$ 
and the effect of uncertainty in $\Sigma_{\rm c}$ is diminishing for DM halos 
with smaller mass\footnote{Note that $Q_{\rm c}$ allows for a smaller variation of order unity.}.
The value of $n_{\rm c}$ is more uncertain and depends largely on numerical resolution. Our adopted
value of $1.0$~cm$^{-3}$ complies with most numerical studies on galactic star formation 
and varying this value
by a factor of 10 can produce only a factor of several variations in $M_{\rm g}^{\rm min}$, owing to
a rather week dependence of the total gas mass on $n_{0,0}$ (see Fig.~\ref{fig3}).

It is worth pointing out that the Kennicutt-Schmidt relation~(\ref{KS})
is based on H$\alpha$ measurements, which only reveal the presence of massive stars in the SF
regions.  Recently, mostly thanks to the GALEX satellite, measurements
of UV fluxes became available for dwarf galaxies.  It turns out
that below $\sim$ 10$^{-2}$ M$_\odot$ yr$^{-1}$, the SF
rate determined by the H$\alpha$ measurements largely underestimates
that based on UV fluxes \citep{Lee09}.  
If the IMF in DGs is top-light (i.e. steeper than the Salpeter slope and with 
a small upper mass cutoff), H$\alpha$ fluxes can be very low even though SF is
active \citep{Pflamm09}. This can affect the threshold value for star formation $\Sigma_{\rm c}$
in equation~(\ref{KS}).

\section{Conclusions}
\label{conclude} 

We have constructed a series of rotating equilibrium galaxies consisting of gas, 
stars, and a fixed DM halo,  with masses of the latter $M_{\rm DM}$  being in the 
$10^7-10^{10}~M_\odot$ range. Our models differ from most previous studies in that we 
self-consistently take into account self-gravity of the gas component. Variations in the gas temperature,
DM halo shape, rotation and non-thermal support, and also in the redshift for galaxy formation 
have been considered. We apply contemporary star formation criteria to the resulting equilibrium 
configurations to estimate the feasibility of large-scale star formation in our models.
For the star formation criteria, we choose the Toomre gravitational stability
criterion with the Toomre parameter smaller than a critical value of $Q_{\rm c}=2.0$ and the 
Kennicutt-Schmidt law with the gas surface density greater than a critical value of
$\Sigma_{\rm crit}=5.0~M_\odot$~pc$^{-2}$. These criteria need to be satisfied simultaneously at least
in some parts of the gas disk in order for the model to be marked as star-formation-allowed (SFA).
In addition, we require that the gas mass 
with number density greater than $n_{\rm }=1.0$~cm$^{-3}$ exceed than $10^{4}~M_\odot$
to allow for a SF event of non-negligible magnitude.

We compare gas masses of the SFA models with the baryonic mass $M_{\rm b}$ derived 
from the $\Lambda$CDM theory and WMAP4 data, for which the ratio of the baryon-to-DM mass 
is 0.17. We find the following:
\begin{itemize}
\item For a given DM halo mass there exists a minimum gas mass $M_{\rm g}^{\rm min}$ that is
needed to achieve a state in which star formation is allowed.  The value of $M_{\rm g}^{\rm min}$
depends crucially on the gas temperature $T_{\rm g}$, the gas spin parameter $\alpha$, the amount
of nonthermal support in the gas disk, and, 
to a somewhat lesser extent, on the redshift for galaxy formation $z_{\rm gf}$. On the other hand,
$M_{\rm g}^{\rm min}$ is rather insensitive to the form of the DM halo and 
to the pre-existing stellar disk, provided that the past SF efficiency does not exceed considerably
0.1.

\item As a rule, $M_{\rm g}^{\rm min}$ is smaller for galaxies with smaller $\alpha$ 
and $T_{\rm g}$ and is greater for objects with greater non-thermal support $\sigma_{\rm eff}$. 
In addition, $M_{\rm g}^{\rm min}$ may be smaller for objects that form at higher redshifts.

\item Depending on the gas temperature $T_{\rm g}$, gas spin parameter $\alpha$, gas 
effective velocity dispersion $\sigma_{\rm eff}$, and the redshift for galaxy formation 
$z_{\rm gf}$, the SFA models may have $M_{\rm g}^{\rm
min}$ that is either greater or smaller than $M_{\rm b}$. Models with 
$M_{\rm DM}\ga 10^{9}~M_\odot$ are usually characterized 
by $M_{\rm g}^{\rm min}\la M_{\rm b}$, implying that star formation in such galaxies is a natural 
outcome of their evolution. 
On the other hand, models with $M_{\rm DM}\la 10^{9}~M_\odot$ are often characterized 
by $M_{\rm g}^{\rm min}\gg M_{\rm b}$, implying
that they need much more gas than available according to the $\Lambda$CDM theory to achieve a state
in which star formation is allowed. 

\item A good agreement of our derived $M_{\rm g}^{\rm min}$ with $M_{\rm b}$ can be achieved 
if the spatially uniform gas temperature  
follows the virial relation $T_{\rm g}\propto M_{\rm DM}^{2/3}$ with a proper choice of the zero point
or some added non-thermal support.
However, the required temperatures for objects with $M_{\rm DM}\la 10^8~M_\odot$ are 
quite low ($\le$ a few hundred Kelvin) and the rotation curves are in poor agreement with 
those observed in DGs.

\item SFA models with $M_{\rm DM}\la 10^{8}~M_\odot$ and $T_{\rm g}\ga 
\mathrm{a~few}\times 10^4$~K have $M_{\rm g}^{\rm min}$ that greatly exceed both 
$M_{\rm b}$ and $M_{\rm DM}$, implying that some star-forming DGs may be baryon-dominated.

\end{itemize}
We find that the gas volume density distribution of our model galaxies is crucially sensitive
to the gas temperature, spin parameter, and redshift of galaxy formation, implying 
a variety of possible equilibrium realizations for objects with the same DM halo mass. 
This means that the evolution of a dwarf galaxy
may follow different pathways after the onset of star formation, depending on the
values of $T_{\rm g}$, $\alpha$, and $z_{\rm gf}$ even for the same DM halo mass.

Our modeling suggests that a star-formation-allowed state is more difficult to achieve in 
DM halos with mass $\la 10^9~M_\odot$ than in their upper-mass counterparts, 
because the required gas mass may be much greater than that
available according to the $\Lambda$CDM theory. This implies that there may be a
critical DM halo mass  below which the likelihood of star 
formation and hence the total stellar mass drops substantially. 
Interestingly enough, the stellar versus DM halo mass relation recently derived by \citet{Guo10}
using the SDSS survey and Millennium Simulations implies the existence of such a threshold value. 
On the other hand, DGs with the gas plus stellar mass greater than that of the DM halo 
are not unheard of and recent observations of the mass-to-light ratios in Virgo Cluster
dwarf ellipticals by \citet{kol11} and in gas-rich DGs by \citet{swat11}
point to the existence of such objects. These observations, along with our results,
suggest that the $\Lambda$CDM paradigm is not universal and significant deviations
from the corresponding $M_{\rm b}$--$M_{\rm DM}$ trend are feasible.

Our study outlines the importance of gas self-gravity (neglected in
practically all hydrodynamical studies of isolated DGs) in
building  equilibrium galaxies. The main argument in favour 
of neglecting the gas self-gravity has been based on the assumption 
that the total gas mass is always much smaller than that of the DM halo.
As we have demonstrated, this assumption may be grossly violated, particularly for low-mass
DM halos with $M_{\rm DM}\le 10^8~M_\odot$. 
The reason is that the baryonic matter has to dominate the dark matter in objects with 
low DM halo masses in order to achieve the Kennicutt-Schmidt SF criterion.
We emphasize that our results are strictly applicable to DM halos that have accumulated their
mass reservoir quasi-statically and remain to be justified for object that have 
undergone a series of violent mergers. At the same time, our main 
conclusions are not affected by moderate perturbations in a quasi-equilibrium state 
and reasonable variations in the adopted values of $\Sigma_{\rm crit}$ and $Q_{\rm c}$.

\section{Acknowledgments} The authors are thankful to the referee for an insightful
report that helped to considerably improve the manuscript. This publication is supported 
by the Austrian Science Fund (FWF).
EV thanks a Lise Meitner Fellowship (Austrian Science Fund FWF), project number M 1255-N16
for financial support. Numerical simulations were done on the
Atlantic Computational Excellence Network (ACEnet) and on the Shared Hierarchical Academic
Research Computing Network (SHARCNET).

\begin{appendix}
\section{Solving for the steady-state equations}


\begin{figure}
  \resizebox{\hsize}{!}{\includegraphics{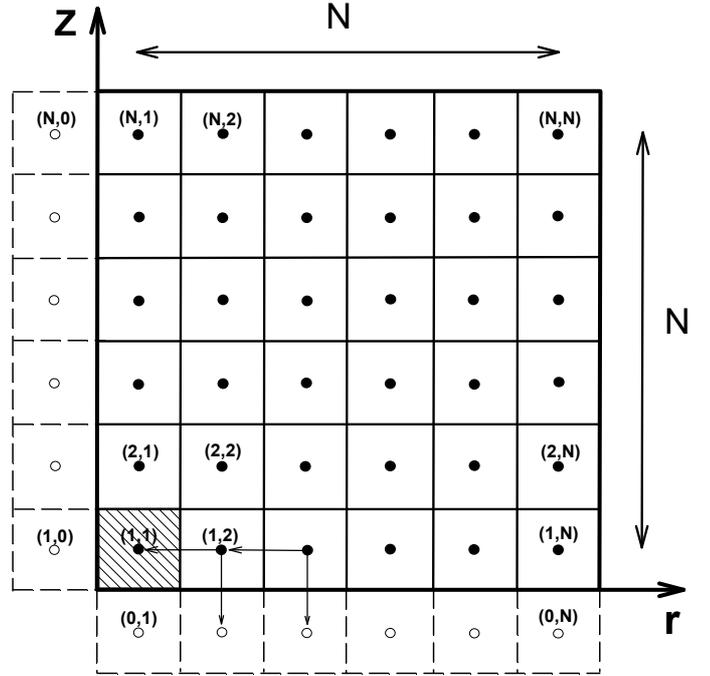}}
      \caption{Computational domain in the $(r,z)$ plane showing active and ghost grid zones with
      solid and dashed lines, respectively. The zone centers are marked with filled/open 
      circles for active/ghost zones. The arrows indicate the backward finite-difference scheme used
      to discretize spatial derivatives. The innermost grid cell highlighted with a backslash palette
      refers to the seed value of the gas surface density $n_{0,0}$.}
         \label{fig1A}
\end{figure}

Figure~\ref{fig1A} shows the $N\times N$ computational mesh employed
to discretize the steady-state equations~(\ref{rhoR}) and (\ref{rhoZ}). 
The active zones are outlined with the solid lines, whereas the two rows of ghost zones 
(representing the reflecting
boundary conditions along the $z$- and $r$-axes) are marked with the dashed lines.
The active/ghost zone centers are denoted with filled/open circles.

A class of problems that does not require the knowledge of the gas density 
at the outer $z$ and $r$ boundaries (i.e., at $N+1$ grid zones) 
can be solved using the following procedure.
We use a first-order backward difference scheme (schematically shown
by the arrows) to  obtain a finite-difference representation of equations~(\ref{rhoR})
and (\ref{rhoZ}) for the case with a spatially uniform $\sigma_{\rm g}$
\begin{eqnarray}
{\ln\rho^{(i,j)}_{\rm g} -\ln\rho_{\rm g}^{(i,j-1)} \over \triangle r_j} &=& 
{1 \over \sigma^2_{\rm g}}
\left[{ v_{\rm g}^{2,(i,j-1/2)} \over r_{j-1/2} } + g_{\rm g,r}^{(i,j-1/2)} + g_{\rm h,r}^{(i,j-1/2)}
\right] \nonumber \\
{\ln\rho_{\rm g}^{(i,j)} -\ln\rho_{\rm g}^{(i-1,j)} 
\over \triangle z_i} &=& {1 \over \sigma^{2}_{\rm g}}\left[ g_{\rm g,z}^{(i-1/2,j)} + g_{\rm h,z}^{(i-1/2,j)} \right],
\end{eqnarray}
where the indices $i$ and $j$ correspond to the $z$ and $r$ coordinate directions, respectively.
We note that densities are defined at the zone centers while gravitational accelerations and velocities
are defined at the corresponding zone interfaces.
In order for this difference scheme to work, one needs to define the gas density at the ghost
zones (which equal to those at the nearest active zones) and also the gas density at the innermost
active zone $(1,1)$ denoted in the paper as the seed density $n_{0,0}$. The corresponding zone is
highlighted with the backslash palette in Figure~\ref{fig1A}. 

With this choice of the discretization scheme and boundary conditions, 
one can notice that $\rho^{(i,j-1)}_{\rm g}$ and  $\rho^{(i-1,j)}_{\rm g}$ are known
for every value of $\rho^{(i,j)}_{\rm g}$ and the latter 
can be found by a fast forward substitution algorithm if one proceeds 
from left to right along the $r$-direction, starting from the bottom layer of zones
and advancing one horizontal layer after another in the direction of increasing $z$. 


%

\section{Testing equilibrium configurations}
\label{tests}
An important reliability check on the solution procedure is to test 
how our equilibrium configurations can be handled
by time-dependent numerical hydrodynamics codes. If our steady-state models
are correct, than a galaxy should stay in rotational equilibrium for
at least 500~Myr, a typical time of interest when simulating the effect of supernova explosions
in DGs.

To perform such a test, we use our time-dependent numerical hydrodynamics code employed earlier
to study the effect of SN explosions in DGs in the local Universe and and large redshifts
\citep{vorob04,VB05,VVS08}. We intentionally turn off cooling and heating to avoid 
the system drifting out of equilibrium due to thermal effects. For the test, we use 
the reference model with $M_{\rm DM}=10^9~M_\odot$, $\alpha=0.9$, $T_{\rm g}=10^4$~K,
and $n_{0,0}=5.0$~cm$^{-3}$. The gas surface density, $Q$ parameter, and velocity profiles
of this model are shown by the red solid lines in the second top row of Figure~\ref{fig2}.

\begin{figure}
  \resizebox{\hsize}{!}{\includegraphics{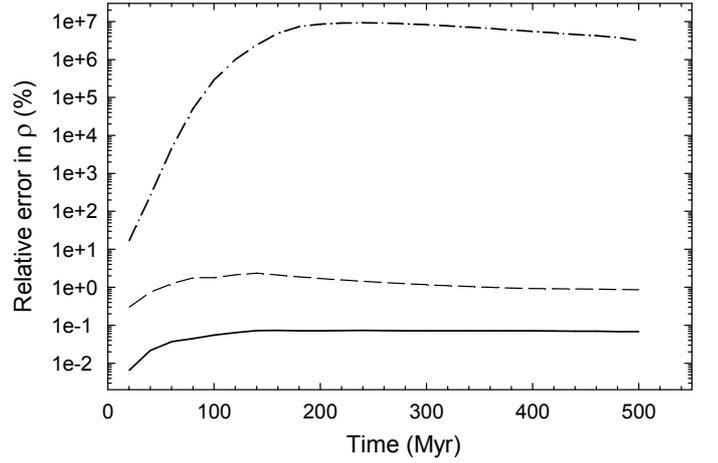}}
      \caption{Maximum and mean relative errors (dashed and solid lines, respectively) in the gas 
      volume density (top) as functions of time $t$
      for the reference test model described in Appendix B. 
      The errors are calculated relative to the initial equilibrium configuration at $t=0$~Myr.
      The dash-dotted lines show the relative errors in the absence of gas self-gravity.}
         \label{fig3A}
\end{figure}

Figure~\ref{fig3A} shows the mean relative error (solid line) and the maximum relative error (dashed
line) in the gas volume density $\rho_{\rm g}$ (top) 
as a function of time $t$ in our test model. 
The relative errors (in per cent) are calculated at every grid cell as\footnote{An extended 
definition of the relative error that takes into account a situation when the gas density declines 
with time, and is thus normalized to the current value of $\rho_{\rm g}(t)$ rather than to the initial
one $\rho_{\rm g}(0)$, yields very similar results.}
\begin{equation}
\triangle \rho_{{\rm g},i,j} =  { | \rho_{{\rm g},i,j}(t) -\rho_{{\rm g},i,j}(0) |
\over \rho_{{\rm g},i,j}(0) }
\end{equation}
and demonstrate the degree to which our equilibrium is held 
by the code during the time evolution. The mean relative errors $\overline{\triangle \rho_{\rm g}}$
are calculated by averaging the individual errors $\triangle \rho_{{\rm g},i,j}$ over the 
entire computational grid.
As one can see, the mean relative errors never exceed 0.1\%,
meaning the equilibrium is well preserved globally. The maximum relative error never exceeds 3\% and
is kept below 1\% during  most of the evolution. We note the maximum errors occur
in dynamically unimportant regions near the axes at large $\varpi$. The gas temperature shows
essentially the same behaviour. This test convincingly proves the robustness and 
reliability of our solution procedure.

To demonstrate the importance of gas self-gravity and to perform the final check on
our self-gravitating equilibrium configurations, we artificially turn off the gas self-gravity
in our time-dependent numerical hydrodynamics code. The dash-dotted lines in Figure~\ref{fig3A} present
the resulted mean relative error. It is obvious that neglecting the gas self-gravity results in a complete
destruction of the equilibrium state, with the mean relative errors exceeding $10^{6}$ for the gas
volume density!

\end{appendix}

\end{document}